\documentclass[acmsmall,nonacm,authorversion,screen]{acmart}

\usepackage{comment}
\usepackage{xspace}

\usepackage{amsmath}
\usepackage{amsthm}
\usepackage{stmaryrd}

\newtheorem{theorem}{Theorem}

\usepackage{fancyvrb}
\usepackage{hyperref}
\usepackage{xcolor,colortbl}
\usepackage{tikz}
\usetikzlibrary{cd}
\usetikzlibrary{matrix}
\usepackage{array}
\usepackage{mathtools}
\usepackage{wrapfig}

\newcommand{\figref}[1]{Figure~\ref{fig:#1}}
\newcommand{\figsref}[2]{Figures~\ref{fig:#1} and~\ref{fig:#2}}

\newcommand{\secref}[1]{Section~\ref{sec:#1}}

\newcommand{\thmref}[1]{Theorem~\ref{thm:#1}}
\newcommand{\defref}[1]{Definition~\ref{def:#1}}

\usepackage{array}

\newcolumntype{L}{>{$}l<{$}} 

\newenvironment{eqnarr} 
  {\[\begin{array}{r@{\;}c@{\;}l}}
  {\end{array}\]}

\def\CB{\color{blue}}
\def\CR{\color{red}}
\def\CG{\color{darkgray}}

\def\CCB{\CN}
\def\CCR{\CN}
\def\CCG{\CN}

\usepackage{listings}
\usepackage{etoolbox}
\expandafter\patchcmd\csname \string\lstinline\endcsname{%
  \leavevmode
  \bgroup
}{%
  \leavevmode
  \ifmmode\hbox\fi
  \bgroup
}{}{%
  \typeout{Patching of \string\lstinline\space failed!}%
}

\let\doublebar\| 
\def\DELKWS{\lstset{keywords=[1]{},keywords=[2]{},keywords=[3]{},keywords=[4]{},keywords=[5]{}}}
\def\|{\lstinline[language=Caml,mathescape,keywords={}]|}

\newcommand{\equaldef}{\stackrel{\scriptscriptstyle\textup{\textsf{def}}}{=}}

\newcommand{\MAP}{\varphi}
\newcommand{\FIX}{\textsc{fix}}
\newcommand{\FV}{\textsc{fv}}
\newcommand{\DV}{\textsc{dv}}

\newcommand{\DATATYPE}{T}
\newcommand{\TUNK}{\DATATYPE_{\UNK}}

\newcommand{\TUNKABS}{\DATATYPE_{\UNK\ABS}}
\newcommand{\TUNKABSERR}{\DATATYPE_{\UNK\ABS\ERR}}

\newcommand{\T}{\textup{\textsf{true}}}
\newcommand{\F}{\textup{\textsf{false}}}
\newcommand{\VP}{\textup{\textsf{?}}}
\newcommand{\UNK}{\bot}
\newcommand{\ABS}{\square}
\newcommand{\ERR}{\mathord{\nshortparallel}}
\newcommand{\EE}{\ERR}

\newcommand{\DKAHN}{D_\textsc{k}}

\newcommand{\DSYNC}{D_\textsc{}}

\newcommand{\CONS}{\cdot}

\newcommand{\LTES}{\sqsubseteq_\textsc{}}

\newcommand{\GTS}{\sqsupset_\textsc{}}

\newcommand{\CUPS}{\sqcup_\textsc{}}

\newcommand{\EI}{i}
\newcommand{\ES}{s}
\newcommand{\EB}{b}
\newcommand{\ET}{t}
\newcommand{\EF}{f}

\newcommand{\LEN}[1]{|#1|}
\newcommand{\STH}[2]{#1_{#2}} 
\newcommand{\SHD}[1]{\STH{#1}{0}} 
\newcommand{\STL}[1]{\STH{#1}{1:}} 

\newcommand{\KF}[1]{\mathop{\DELKWS#1^{\#}_{\textsc{k}}}}

\newcommand{\SF}[2][.]{\mathop{\DELKWS#2^{\#}_{\textsc{}\if#1.\relax\else#1\fi}}}
\newcommand{\PRESENTNEXT}{\|present_next|}

\newcommand{\VENV}[1][.]{V\if#1.\relax\else_{#1}\fi}

\newcommand{\BIND}{\mathop{:=}}

\newcommand{\TJ}[1][]{{#1}\vdash} 

\newcommand{\SEM}[1][]{{\mkern-.5mu}^{\phantom{\#\hskip-1ex}}_{\textsc{#1}\!}\llbracket}
\newcommand{\KSEM}[1][.]{\SEM[k]\if#1.\relax\else\TJ[#1]\fi}
\newcommand{\ASEM}[1][.]{\SEM[a]\if#1.\relax\else\TJ[#1]\fi}
\newcommand{\SSEM}[1][.]{\SEM[]\if#1.\relax\else\TJ[#1]\fi}
\makeatletter
\def\ENDSEM{%
  \@ifnextchar[%
    {\ENDSEM@i}
    {\ENDSEM@i[.]}%
}
\def\ENDSEM@i[#1]{%
  \@ifnextchar[%
    {\ENDSEM@ii{#1}}
    {\ENDSEM@ii{#1}[.]}%
}
\def\ENDSEM@ii#1[#2]{%
  \rrbracket\if#1.\relax\else^{\if#2.\phantom{\#\hskip-1ex}\else\vec{}#2\fi}_{#1}\fi
}
\makeatother

\newcommand{\DSEM}{\llbracket}

\newcommand{\ASSIGN}[4]{{#1}[{#2}\leftarrow_{#3}{#4}]}

\newcommand{\WORKLIST}{\textit{worklist}}

\newcommand\II{\overline{\imath}}

\newcommand{\BISEM}[1]{\ensuremath{\DSEM {#1} \ENDSEM}}
\newcommand{\BIFIX}[2]{\ensuremath{\FIX({#1},{#2})}}

\newcommand{\RESTRICT}[2]{\ensuremath{{#1}\mid_{#2}}}
\newcommand{\EXTEND}[2]{\ensuremath{{#1}\upharpoonright_{#2}}}

\newcommand{\ITER}[1]{\{\!\!\{{#1}\}\!\!\}}
\newcommand{\SITER}[1]{\{\!\!\{{#1}\}\!\!\}_\#}
\newcommand{\VAL}[2]{\ensuremath{{#1}[{#2}]}}
\newcommand{\AASSIGN}[3]{\ensuremath{{#1}[{#2}\twoheadleftarrow{#3}]}}
\newcommand{\AITER}[2]{\{\!\!\{\!\!\{{#1}\}\!\!\}\!\!\}_{#2}}

\newcommand{\REACT}{\ensuremath{\textrm{R}(\textrm{ML})^2}\xspace}

\begin{document}

\title{Bidirectional Reactive Programming for Machine Learning}

\author{Dumitru Potop Butucaru}
\affiliation{%
  \institution{Inria}
  \country{France}}
\email{dumitru.potop@inria.fr}

\author{Albert Cohen}
 \affiliation{%
   \institution{Google DeepMind}
   \country{France}}

\author{Gordon Plotkin}
\affiliation{%
  \institution{Google DeepMind}
  \country{USA}}

\author{Hugo Pompougnac}
\affiliation{%
  \institution{Inria}
  \country{France}}

\begin{abstract}
  Reactive languages are dedicated to the programming of systems which interact continuously and concurrently with their environment.
  Values take the form of unbounded streams modeling the (discrete) passing of time or the sequence of concurrent interactions.
  While conventional reactivity models recurrences forward in time, we introduce a symmetric reactive construct enabling backward recurrences.
  Constraints on the latter allow to make the implementation practical.
  Machine Learning (ML) systems provide numerous motivations for all of this:
  we demonstrate that reverse-mode automatic differentiation, backpropagation, batch normalization, bidirectional recurrent neural networks, training and reinforcement learning algorithms, are all naturally captured as bidirectional reactive programs.
\end{abstract}

\maketitle

\section{Introduction}

Conventional machine learning (ML) frameworks offer a tensor-centric view of the design and implementation of deep neural networks.
But ML models do not stand by themselves as pure tensor functions.
ML applications typically \emph{interact with an environment and often operate on streams of data} collected and processed over time. For instance, the reactive control of a self-driving car \cite{tesla,self-driving-planning} operates on streams of data coming from sensors and controlling actuators.
Training algorithms themselves embed a model into a reactive loop: it is decomposed into epochs and (mini-)batches allowing the efficient scheduling of computations and I/O, parameter updates, etc.
The same applies to reinforcement learning (RL) agents \cite{Dul21}.
Back to the automated driving example, {\em stateful} behavior is essential to taking into account previously-inferred facts such as speed limits, whether the current lane is a left turn etc., long after the acquisition of sensor inputs.
Other examples of ML components embedded into stateful reactive feedback loops include model-predictive maintenance \cite{serradilla2020deep}, control \cite{camacho2013model}, and digital twins \cite{doi:10.2514/6.2020-0418}.
ML models themselves involve stateful constructs in the form of recurrent neural network (RNN) layers.
When generating optimized code, even matrix products and convolutions in feedforward networks can be folded over time, using (stateful) buffering to reduce memory footprint \cite{kws-streaming2}.
In distributed settings, the efficient implementation of large models involves pipelined communications and computations \cite{huang2019gpipe,liu2022dive,collectives}, which amounts to locally recovering a streaming execution pattern.

Considering this broad range of scenarios, we observe that existing ML frameworks inadequately capture reactive aspects, raising barriers between differentiable models and the associated control, optimization, and input/output code.
These barriers worsen the gap between ML research and system capabilities \cite{10.1145/3317550.3321441}, particularly in the area of control automation where embedded ML engineering relies on undisclosed, ad-hoc implementations \cite{tesla}.
Our contributions are threefold:
\vspace{-3mm}
\begin{changemargin}{2mm}{0mm}
\item\hspace*{-2mm}{\bf Dataflow reactive machine learning.} We first observe that conventional (dataflow or \texttt{fold\_left}) reactive programming allows the natural specification of a variety of ML inference models, optimization and RL algorithms.
  By indexing data and parameter streams along the logical time steps of computation and interaction, one may integrate the internal processing of ML models with data pre-/post-processing stages, and embed everything into a reactive control loop.
\item\hspace*{-2mm}{\bf Bidirectional reactive programming.} We extend such a core reactive language with recurrences backward in time (\texttt{fold\_right}).
  While classical reactive programs compute the values of cycle $n+1$ from those of cycle $n$, forming recurrences forward in time, bidirectional reactive programs allow a value of cycle $n$ to depend on one of cycle $n+1$.\footnote{``Bidirectional'' borrows here from machine learning terminology. It is not about computing
  from updated outputs \cite{Bidirectional}.}
  This extension captures reverse-mode automatic differentiation, backpropagation through recurrent layers, training parameter update, distributed and pipelined training, bidirectional recurrent networks, and layers crossing data and time boundaries such as batch normalization; all of this without the need for an explicit gradient tape \cite{Rad23} or caching mechanism \cite{Mos20}.
  This extension preserves the bounded time and memory guarantees of reactive languages whenever recurrences backwards in time have a bounded horizon.
\item\hspace*{-2mm}{\bf Formal semantics.} The semantics of this bidirectional reactive language is defined from purely functional stream denotations.
  We provide a denotational semantics reminiscent of synchronous Kahn networks \cite{SKN}.
  To ensure productivity (i.e.\ the absence of deadlocks), chains of backward dependences must \emph{always eventually} terminate, so that execution can be implemented as \emph{globally forwards and locally backwards} with finite buffering.
\end{changemargin}
We unify independently developed ML models and algorithms on top of reactive constructs with deterministic formal semantics. This allows homogeneous and unambiguous specification, spanning from individual layers to RL algorithms. It also paves the way for the combination of efficient ML processing with the scheduling and resource allocation expressiveness of reactive languages.

\section{Related work}

This paper contributes to programming languages for both machine learning and reactive control systems. We address them separately.

\paragraph{Machine learning}
While dataflow modeling has long been the norm in deep learning frameworks, the focus has typically been on efficient data processing, one input (batch) and one version of the trained parameters at a time.
The design of ML compilers has been primarily driven by this common-case scenario, where both inference and training involve computations over data sets that have been already processed and stored.
This contrasts with the needs of control applications that involve reactive and stateful interactions with an environment.
Such interactions are represented outside of the ML model, which makes it difficult to provide implementations that satisfy stringent real-time and memory requirements \cite{tesla,kws-streaming2,video-seg,ml-streaming,self-driving-planning}.

\definecolor{myyellow}{rgb}{1.0, 1.0, 0.8}
\definecolor{myorange}{rgb}{1.0, 0.65,0.0}
\begin{figure}[h!tb]
  \vspace*{-3mm}
  \begin{center}
    {\footnotesize
\begin{tabular}{|l||l|l|l|}
  \hline
                                 & \multicolumn{3}{c|}{Application scenario}                                                         \\
                                 \cline{2-4}
    Network Type                 & Inference                        & Training                         & Reinforcement Learning \\
  \hline
  \hline
\cellcolor{myyellow}Feedforward w/o BatchNorm &\cellcolor{myyellow}Stateless     &\cellcolor{yellow}Stateful, Forward      &\cellcolor{yellow}Stateful, Forward\\
  \hline
\cellcolor{myyellow}Feedforward w/ BatchNorm  &\cellcolor{myyellow}Stateless     &\cellcolor{myorange}Stateful, Bidirectional     &\cellcolor{myorange}Stateful, Bidirectional\\
  \hline
\cellcolor{yellow}Recurrent      &\cellcolor{yellow}Stateful, Forward       &\cellcolor{myorange}Stateful, Bidirectional     &\cellcolor{myorange}Stateful, Bidirectional \\
  \hline
\cellcolor{myorange}Bidirectional&\cellcolor{myorange}Stateful, Bidirectional    &\cellcolor{myorange}Stateful, Bidirectional     &\cellcolor{myorange}Stateful, Bidirectional \\
  \hline
\end{tabular}}
  \end{center}
  \vspace*{-2mm}
  \caption{Expressiveness requirements for neural networks operating on streams w.r.t.\ network type.
    The core language covers the pale and bright yellow cells.
    Bidirectional reactivity is required for the orange cells.}
  \label{fig:complexityclasses}
  \vspace*{-3mm}
\end{figure}

By comparison, we provide a language allowing the natural representation of reactive programs on streams, to express transformations over these such as automatic differentiation.
It has been shown recently \cite{hipeac22} that dataflow reactive programming \cite{lustreRTSS,pouzetLustre,from-simu,endochrony} captures a large class of ML applications---the light and yellow cells of \figref{complexityclasses}.
In particular, the variety of ML constructs in these applications can be reduced to just four dataflow constructs.
Bidirectional reactivity---the orange cells of \figref{complexityclasses}---improves expressiveness further, with support for the automatic differentiation of arbitrary networks and context-dependent operations such as batch normalization (whose semantics differ in inference and training modes).
One no longer needs a different programming model for training in the data center vs.\ inference on the edge, for latency- vs.\ throughput-optimization, or a specific ``out-of-the-language'' toolkit for RL.

\paragraph{Reactive languages}
We aim to facilitate the design and implementation of ML applications by exploiting abstractions and methods originating from reactive systems.
Our main influence is the dataflow synchronous language Lustre, and its commercial version Scade, introduced in the 1980s \cite{lustreRTSS} to allow the formal specification, the efficient execution and the automatic verification of deterministic concurrent systems with feedback and conditional control.
Lustre, as well as the broader class of functional reactive languages \cite{ElliottHudak97:Fran},\footnote{See also \url{https://en.wikipedia.org/wiki/Functional_reactive_programming}.} do not allow the representation of bidirectional recurrences.

Lucid \cite{10.1145/359636.359715}, one of the inspiration for Lustre, did introduce a \texttt{next} construct with semantics similar to the \|post| construct we introduce for bidirectional recurrences.
The designers of Lucid explicitely disallowed recurrences into the future however, limiting the usage of \texttt{next} in right-hand side expressions to non-recursive definitions.

\section{Language syntax and overview}
\label{sec:syntax}

\figref{syntax} introduces the syntax of \REACT---for \underline{R}eactive \underline{M}achine \underline{L}earning \underline{M}eta \underline{L}anguage.
It is based on that of the classical dataflow reactive language Lustre \cite{lustreRTSS,pouzetLustre} from which it borrows its core set of dataflow primitives (in black) \cite{velus}.
The Kleene star stands for comma-separated lists when applied to expressions and variables,
for lists of equations in nodes, and for list of nodes at the top level of the program.
The \|main| node serves as the entry point to the computation; its inputs and outputs characterize the semantics of the whole program.

\begin{figure}[h!tb]
\def\CCB{\CB}
\def\CCR{\CR}
\def\CCG{\CG}
\vspace*{-2mm}
{\small%
$$
\begin{array}{rrlL}
  e & ::= & x \;|\; c & variable, constant stream \\
    &   | & f ( e^\ast ) & node instantiation and function application \\
    &   | & e \; \lstinline|when| \; e & sub-sampling (for conditional execution) \\
    &   | & \lstinline|merge| \; e \; e \; e & merging (combining, up-sampling) \\
    &   | & e \; \lstinline|fby| \; e & recurrence forward in time (initialized delay) \\
    &   | & \lstinline|post| \; e  & recurrence backward in time \\
  \textit{eq} & ::= & \textit{x}^\ast \; \lstinline|=| \; e \; \lstinline|;| & stream equation \\
 \textit{node} & ::= & \lstinline|node| \; f \lstinline|(| \textit{x}^\ast \lstinline|)->(| \textit{x}^\ast \lstinline|)| \; \textit{eq}^\ast & node \\
 \textit{prog} & ::= & \textit{node}^\ast & program
\end{array}
$$}
\vspace*{-5mm}
\caption{Syntax of the \REACT reactive language.}
\vspace*{-3mm}
\label{fig:syntax}
\end{figure}

To this classical reactive core, \REACT adds the single new primitive \|post|.
It brings a fundamental extension over classical reactive languages: recurrences backwards in time, i.e. the ability to traverse time backwards.
In a ML context, this is required to represent bidirectional networks \cite{brnn}, the training or reinforcement learning of stateful networks by means of backpropagation through time \cite{fstbptt,bptt}, and batch normalization \cite{batchNorm}.
The expressive power of this construct will be discussed in \secref{revtime} and its formal semantics will be defined in \secref{semantics}.

Note that \REACT features neither general recursion nor higher order functions.
This choice, inherited from the synchronous reactive language Lustre, allows implementation in bounded time and memory.\footnote{E.g.\ for embedded and real-time applications.}
More expressive variants of the language could be considered in the future, in the spirit of FRP \cite{ElliottHudak97:Fran} or Reactive ML \cite{DBLP:conf/ppdp/MandelP05}, but this is left for future work.

For brevity, we did not include tensor notation and tensor algebra operations in this syntax.
A few of the examples below use conventional Python notations and Numpy or Keras operations, with square brackets for functional arrays.

As noted in \figref{complexityclasses} and justified in the next section, the core language (i.e.\ without \|post|) allows the natural representation of inference and reinforcement learning, for feedforward networks and one-directional RNNswithout batch normalization.
This includes models with complex control such as gated mixture of experts \cite{sparsely-gated}.
In the case of feedforward networks, it can also represent forward- and reverse-mode differentiation, backpropagation, a wide range of (non-pipelined) training algorithms, and mixing ML models with more general reactive control.
We dedicate \secref{core} to presenting its expressive power from an ML perspective.

Note that synchronous reactive languages like Lustre have \emph{clocks}, a dependent type system making explicit the otherwise implicit wellformedness of sampled stream expressions, and the associated conditional execution \cite{clocks}.
The inability to infer the clock of a variable results in the program being rejected as incorrect, while properly clocked reactive programs are called \emph{synchronous} \cite{lustreRTSS}.
We leave clocks out of this paper for the sake of conciseness, since for the most part, applications to ML and bidirectional extensions are orthogonal to clock types and inference.

\section{Dataflow reactive machine learning}
\label{sec:core}

\paragraph{Informal semantic principles}
The underlying reactive language without \|post| allows the definition of (forward) dataflow \emph{nodes} that compute \emph{streams} of outputs from streams of
inputs.
It has a concurrent semantics that parallels that of the classical \|fold_left| iterator in functional languages.
While classical functional languages evaluate recurrences \emph{in space}, left-to-right along lists or vectors, the iterations of a reactive node unfold \emph{in time} along potentially infinite input streams of data.
At each cycle, the inputs of one recurrence step are acquired along with the current value of the \|fold_left| accumulator, known as the \emph{program state} in reactive programming; then computations for this cycle take place, yield the associated outputs and the new program state (the value of the accumulator), the latter are propagated to the next cycle before its evaluation may start.

The motivations for choosing a concurrent dataflow semantics are similar in both reactive control \cite{lustreRTSS} and ML applications \cite{abadi-tf-semantics}.
It captures the intrinsic concurrency of sensors and actuators in control automation, and the intrinsic concurrency of compute graphs in ML; it natively models parallel and distributed execution, including pipelining \cite{whitlock-tf-semantics}, and it provides ample freedom to the compiler to implement graph-level optimizations such as different forms of fusion \cite{XLA}.

Furthermore, a reactive language generalizes the \|fold_left| iterator by allowing conditional input/output.
One cycle does not need to acquire a value on every input stream and produce a value on every output stream. Instead, any variable can be marked as \emph{absent} in a cycle.
Within each cycle, execution is data-driven: computations take place as soon as enough input variables have acquired a value.
As a result, \emph{variable uses need not be dominated by their definition}: the syntactic ordering of equations in a node is semantically irrelevant.

\paragraph{Core constructs}
The reactive core of \REACT uses only four dataflow primitives to implement these semantic principles: function application, initialized delay, conditional (sub-)sampling, and conditional up-sampling a.k.a.\ merging.
All other constructs of Lustre can be reduced to the four primitives by structural expansion, and thus can be implemented on top of core \REACT. 
The usual scalar operators (\|+|, \|*|, etc.), lifted to tensors through point-wise application, and other pure functions imported from a classical (non-reactive) language can be used in \REACT programs. The function application primitive of \REACT lifts them to streams by pointwise application at each time step. Among these pure functions we mention the (eager) conditional expression on streams $\|if| \; e_1 \; \|then| \; e_2 \; \|else| \; e_3$, which does not need special handling as a primitive.

The remainder of this section explores these constructs and their intuitive semantics in greater detail.
Much of the discussion is based on the example of \figref{gdescent}, which implements the gradient descent of a two-layer feedforward network.
To simplify the presentation, we apply the layers to scalars instead of higher-dimensional tensors, and we omit the batch dimension.
The backpropagation control structure is not simplified however.
The model has two parameters (in layer \|dense|); it exercises all \REACT primitives, including forward recurrences---in blue---and backward recurrences---in red---and as we shall see, it already goes beyond the ML state of the art.

\lstset{
  numbers=left,
  language=[Objective]{Caml},
  keepspaces,
  columns=flexible,
  mathescape,
  basicstyle=\ttfamily,
  commentstyle=\upshape,
  keywordstyle=[1]{\bfseries},
  keywordstyle=[2]{\CCB\bfseries},
  keywordstyle=[3]{\CCR\bfseries},
  keywordstyle=[4]{\CCG},
  keywordstyle=[5]{\underbar},
  otherkeywords={::},
  keywords=[1]{node,when,merge},
  keywords=[2]{fby,pre,restart,every},
  keywords=[3]{post,rev},
  keywords=[4]{on,::},
  keywords=[5]{param}
}

\begin{figure}[h!tb]
\vspace*{-4mm}
\begin{center}
  \begin{minipage}{12cm}
\def\CCB{\CB}
\def\CCR{\CR}
\def\CCG{\CG}
\begin{lstlisting}[basicstyle=\ttfamily\footnotesize,lineskip=-2pt]
node param (init)->(o)                 node diff_param(init,bp,do)->(o)
  o = init;                              o = init fby o1;
                                         o1 = o + merge bp do 0.0;
node dense(i)->(o)                     node diff_dense(i,bp,do)->(o,di)
  b = param(0.0);                        b = diff_param(0.0,bp,db);
  k = param(1.0);                        k = diff_param(1.0,bp,dk);
  o = k * i + b;                         o = k * i + b;
                                         di = k * do;               (* $\partial o/\partial i=k$ *)
                                         dk = i * do;               (* $\partial o/\partial k=i$ *)
                                         db = do;                   (* $\partial o/\partial b=1$ *)
node multiply(i)->(o)                  node diff_multiply(i,bp,do)->(o,di)
  o = i * i;                             o = i * i;
                                         di = do * 2.0 * i when bp; (* $\partial o/\partial i = 2i$ *)
node app(i)->(o)                       node diff_app(i,bp,gt,learn_rate)->(o)
  x = dense(i);                          x,di = diff_dense(i,bp,dx);
  o = multiply(x);                       o,dx = diff_multiply(x,bp,do);
                                         do = -learn_rate * mse_grad_i(o when bp,gt);
node app_pipe(i)->(o)                  node diff_app_pipe(i,bp,gt,learn_rate)->(o)
  x = dense(i);                          x,di = diff_dense(i,bpi,dx);
  y = 0.0 fby x;                         y,bpi,dx = diff_fby1(0.0,x,bp,dy);
  o = multiply(y);                       o,dy = diff_multiply(y,bp,do);
                                         do = -learn_rate * mse_grad_i(o when bp,gt);
                                       node diff_fby1(init,i,bpo,do)->(o,bpi,di)
                                         o = init fby i;
node mse_grad_i(i,gt)->(o)               bpi = post bpo;
  o = 2.0 * (i-gt);                      s = merge bpo do ((post s) when not bpo);
                                         di = s when bpi;
node infer(i)->(o)                     node train(i,gt)->()
  o = diff_app_pipe(i,false,0.0,0.0);    _ = diff_app_pipe(i,true,gt,0.01);
\end{lstlisting}
  \end{minipage}
  \vspace*{-4mm}  
\end{center}
  \caption{Complete gradient descent implementation in \REACT. On the
    left (lines 1-21) the specification (node \texttt{app}) and a pipelined
    version of it (node \texttt{app\_pipe}).
    On the right, the gradient descent implementation for
    every node, obtained by 
    structural translation not detailed in this paper. At the bottom,
    inference and training nodes obtained by specialization of the
    gradient descent node \texttt{diff\_app\_pipe}.}
  \label{fig:gdescent}
\vspace*{-3mm}
\end{figure}

The two-layer feedforward network implements the function
$\texttt{app}(\texttt{i}) = \texttt{multiply}(\texttt{dense}(\texttt{i})) = (\texttt{k} \times \texttt{i} + \texttt{b})^2$, where
$\texttt{k}$ and $\texttt{b}$ are parameters initialized with values $1.0$ and $0.0$, respectively. During training, the objective is to
perform gradient descent on the parameters in order to reduce the Mean Square Error (MSE) with respect to a
ground truth \texttt{gt}. If the ground truth provided for input \texttt{i} is $\texttt{gt}(\texttt{i})=(2.0 \times \texttt{i} - 3.0)^2$,
the parameters will converge towards 2.0 and -3.0, respectively.

\subsection{Hierarchy, dataflow equations, normalization}

An \REACT program consists of a list of dataflow \emph{nodes}. Each
node comprises an \emph{interface} and zero or more \emph{equations}
connected into a dataflow graph by means of variables. In an ML
context, \REACT dataflow nodes represent elementary computations,
layers, inference, training, or the full application.
A dataflow node such as \|app| can {\em instantiate} another. 
Semantically, dataflow instantiation (which we inherit from Lustre) has the same effect as inlining the code of the instantiated node.\footnote{The \|param| node deserves a specific note: it has dedicated, non-standard translation during the synthesis of backpropagation code, as discussed in \secref{backprop}.  Yet in inference mode \|param| translates to the identity function.  For consistency throughout this paper, we assume that the synthesis of backpropagation code is performed prior to any inlining of \|param|.}
Due to this inlining semantics, while code may be reused across multiple instances of the same node, the storage of stateful nodes is \emph{not} shared between instances. In our example, node \|diff_dense| contains two instances of node \|diff_param|, each one independently storing the value of one parameter during training (using the \|fby| primitive at Line~2).

Dataflow dependences between primitives of a node are either explicitly represented using named variables, or implicitly within expressions. For instance, in node \|dense|, in Line~8, the operator ``\texttt{+}'' takes as input the output of operator ``\texttt{*}''.
It is always possible to transform equations with complex expressions into multiple single-primitive equations by exposing implicit dataflow dependences with fresh variables. This process is traditionally referred to as {\em normalization} and is analogous to the administrative normal form in continuation-passing style.
The result of this process for node \|app|, combined with the inlining of the instantiation of \|dense| and \|multiply| (for easier reference in simulations), is provided in \figref{norm}.
Note in lines~4-5 the single-primitive equations produced as part of the normalization of the dataflow expression in Line~8 of node \|dense|.

\subsection{Function application}

As explained earlier, at each execution cycle, an equation performs its computation as soon as its variables have acquired a value are are known as absent.

\tikzset{ 
    table/.style={
        matrix of nodes,
        nodes={
            rectangle,
            draw=black,
            align=center
        },
        minimum height=1.0em,
        text depth=0.5ex,
        text height=1.0ex,
        text width=3em,
        nodes in empty cells,
        column 1/.style={
          nodes={text width=2em,font=\bfseries}
        },
    }
}
\begin{figure}[h!tb]
  \begin{minipage}{.35\textwidth}
\vspace*{-2mm}
\begin{lstlisting}[basicstyle=\ttfamily\footnotesize,lineskip=-2pt]]
  node app(i)->(o)
    b = 0.0;
    k = 1.0;
    t = k * i;
    x = t + b;
    o = x * x;
\end{lstlisting}
\vspace*{-4mm}
\caption{Inlining and normalization of the \|app| node}
\label{fig:norm}
\vspace*{-4mm}
  \end{minipage}
  \hfill
  \begin{minipage}{.55\textwidth}
\vspace*{-4mm}
    \footnotesize
\begin{tikzpicture}
\matrix (first) [table]
{
cycle & 0               & 1                & 2                & 3                & 4               \\
t     & \node(y0){$1$}; &  \node(y1){$2$}; &  \node(y2){$1.5$}; &  \node(y3){\CG$\ABS$}; & \node(y4){$1$}; \\
b     & \node(y0){$0$}; &  \node(y1){$0$}; &  \node(y2){$0$}; &  \node(y3){\CG$\ABS$}; & \node(y4){$0$}; \\
x     & \node(y0){$1$}; &  \node(y1){$2$}; &  \node(y2){$1.5$}; &  \node(y3){\CG$\ABS$}; & \node(y4){$1$}; \\
o     & \node(z0){$1$}; &  \node(z1){$4$}; &  \node(z2){$2.25$}; &  \node(z3){\CG$\ABS$}; & \node(z4){$1$}; \\
};
\draw [->,brown,thick] (-1.4,        0.4) -- (-1.4,       -0.4);
\draw [->,brown,thick] (-1.4+  1.05, 0.4) -- (-1.4+  1.05,-0.4);
\draw [->,brown,thick] (-1.4+2*1.05, 0.4) -- (-1.4+2*1.05,-0.4);
\draw [->,brown,thick] (-1.4+4*1.05, 0.4) -- (-1.4+4*1.05,-0.4);
\draw [->,brown,thick] (-2.0,       -0.1) -- (-2.0,       -0.4);
\draw [->,brown,thick] (-2.0+  1.05,-0.1) -- (-2.0+  1.05,-0.4);
\draw [->,brown,thick] (-2.0+2*1.05,-0.1) -- (-2.0+2*1.05,-0.4);
\draw [->,brown,thick] (-2.0+4*1.05,-0.1) -- (-2.0+4*1.05,-0.4);
\draw [->,brown,thick] (-2.0,       -0.5) -- (-2.0,       -0.8);
\draw [->,brown,thick] (-2.0+  1.05,-0.5) -- (-2.0+  1.05,-0.8);
\draw [->,brown,thick] (-2.0+2*1.05,-0.5) -- (-2.0+2*1.05,-0.8);
\draw [->,brown,thick] (-2.0+4*1.05,-0.5) -- (-2.0+4*1.05,-0.8);
\end{tikzpicture}
\vspace*{-3mm}
    \caption{Sample chronogram of the equations in lines~5-6 of \figref{norm}. Brown arrows represent data dependences.}
    \label{fig:chronogramapply}
    \vspace*{-4mm}
  \end{minipage}
\end{figure}

The function application primitive of \REACT incorporates classical
(non-reactive, non-streaming) functions into our reactive language by
means of pointwise application. At each cycle, the function
application primitive will wait for one value on each input variable
before calling the classical function and then assigning its results
to the output variables. If one input or output is absent during the
cycle, then all inputs and outputs must be absent and the function
is not called. If we denote with $t_0<t_1<t_2<\ldots$
the indices of the cycles where the inputs and outputs of the
function application
``$\|o|_1,\ldots,\|o|_m \; = \; \|f|(\|i|_1,\ldots,\|i|_n)$'',
are present and with $y_n$ the value of a variable $y$ at cycle $n$, then
we will have, for all $n\geq 0$:
$\|o|_{1_{t_n}},\ldots,\|o|_{m_{t_n}} \; = \; \|f|(\|i|_{1_{t_n}},\ldots,\|i|_{n_{t_n}})$.
We depict in \figref{chronogramapply} a possible execution trace---known as a \emph{chronogram}\,\footnote{A chronogram displays the successive values of streams of interest at every cycle index in a given range. Cyclic execution allows us to graphically align multiple streams, since indexes coincide with computation cycles.}---of the equations in lines~5-6 of \figref{norm}.
The trace is 5 cycles long (indices~0 to~4).
In cycle~3 all variables are absent---the {\CG$\ABS$} symbol.

Notice how, inside each cycle, the computation is \emph{causal}, with
the producer of a variable value being executed before all consumers.
Function applications can be composed into arbitrarily complex
(hierarchic) stateless computations, which can represent the
computation of layers (such as \|dense| in \figref{gdescent}), or full
feedforward models, such as \|app|. The main constraint here is that
\emph{the resulting dataflow graphs are acyclic}.\footnote{In a
cyclic graph formed only of function applications, a computation would
depend on its own input, resulting in deadlock.}

Most functions used in ML applications are pure functions such as
additions, convolutions, \|tanh|, \|sigmoid|. However, ML
practice also makes significant use of functions that are not
pure. These are random functions used in initializers
(e.g.\ \|glorot_uniform|, \|orthogonal|) or in \|dropout| layers.

\subsection{Stateful computations}

We have seen in the previous section how function applications can be
composed into arbitrarily complex stateless computations represented
with acyclic dataflow graphs. However, {\em ML practice always
  involves stateful behaviors} implementing forward recurrences along
the time direction. As we shall see, these recurrences are mandatory
in the training process (to store the parameter values), but they may
already appear at specification level, as in recurrent neural networks
(RNNs). Beyond the ML code itself, ML-based systems design also
involves significant stateful components, such as data pre-processing
involving sliding windows\,\footnote{Also known as rotating buffers.} or
efficiently-scheduled pipelined implementations
\cite{huang2019gpipe,whitlock-tf-semantics}.  The \REACT language
construct and a few applications are discussed in this section, while
more advanced uses of stateful behaviors in ML are discussed later, in
\secref{advanced}.

\subsubsection{Initialized delay} 

\begin{wrapfigure}{r}{0.6\textwidth}
  \begin{center}
    {\footnotesize
\vspace*{-3mm}    
\begin{tikzpicture}
    [node distance=1cm]     
\matrix (first) [table]
{
cycle & 0                 & 1                  & 2                 & 3                  & 4                & 5  \\
i     & \node(z0){$3.0$}; &  \node(z1){$5.1$}; &  \node(z2){\CG$\ABS$}; &  \node(z3){$6.1$}; & \node(z4){$3.0$}; & \node(z5){$2.2$};    \\
x     & \node(z0){$4.3$}; &  \node(z1){$0.8$}; &  \node(z2){\CG$\ABS$}; &  \node(z3){$3.3$}; & \node(z4){$1.9$}; & \node(z5){$7.7$};    \\
y     & \node(x0){$3.0$}; &  \node(x1){$4.3$}; &  \node(x2){\CG$\ABS$}; &  \node(x3){$0.8$}; & \node(x4){$3.3$}; & \node(x5){$1.9$};   \\
};
\draw [->,brown,thick] (-2.5,        0.1) -- (-2.5,       -0.6);
\draw [->,blue,thick]  (-1.9,       -0.3) -- (-1.4,       -0.6);
\draw [->,blue,thick]  (-1.9+  1.05,-0.3) -- (-1.4+2*1.05,-0.6);
\draw [->,blue,thick]  (-1.9+3*1.05,-0.3) -- (-1.4+3*1.05,-0.6);
\draw [->,blue,thick]  (-1.9+4*1.05,-0.3) -- (-1.4+4*1.05,-0.6);
\end{tikzpicture}}
\vspace*{-3mm}    
    \caption{Chronogram of \|y = i fby x|. Forward dependences in blue. In brown, the dependence of \|y| on \|i| in the first cycle}
    \label{fig:chronogramfby}
\vspace*{-4mm}    
\end{center}
  \end{wrapfigure}
The \REACT and Lustre statement allowing to cover all these needs is
\|fby|.
Its general form is $y = i \; \|fby| \; e;$ where $i$ and $e$ are expressions of the same type and $y$ is a stream variable.
Its semantics is that of a loop-carried dependence, and it requires that $i$, $e$ and $y$ are all present at the same cycles.
We will denote with $t_0<t_1<t_2<\ldots$ the indices where $i$, $x$ and $y$ are all present.
Then, $y_{t_0}= i_{t_0}$ and $y_{t_n}=e_{t_{n-1}}$ if $n>0$.
\figref{chronogramfby} shows one possible execution trace for a \|fby| equation.

Notice that, unlike function application, the output of
\|fby| does not depend {\em within the current cycle} on its second input. This allows
the construction of cyclic dataflow graphs, such as the one in 
\begin{wrapfigure}{l}{0.3\textwidth}
\def\CCB{\CB}
\def\CCR{\CR}
\def\CCG{\CG}
\vspace*{-3mm} 
\begin{lstlisting}[basicstyle=\ttfamily\footnotesize,lineskip=-2pt]
  node counter()->(o)
    o = 0 fby u;
    u = o + 1; 
\end{lstlisting}
\vspace*{-3mm}
\caption{Counter node in MLR}
\label{fig:counter}
\vspace*{-3mm}
\end{wrapfigure}
\noindent \figref{counter}. In the first cycle (of index
0) its output (the value of variable \|o|) is the value of the first
input, which is 0.  In cycle 1, it is the value of its second input
(variable \|u|) in cycle 0, which is 1. Inductively, we can prove that
in every cycle $n$ the value of \|o| is $n$.

Note that each \|fby| equation of a node defines one element
of the recurrence accumulator which defines the program state.
Thus, the \|counter| node above has a state formed of one integer,
whereas the state of node \|app| in \figref{norm} is formed of
two floating-point values.
For a node $n$ that has not yet been inlined, the state is formed of the
\|fby| equations defined directly in $n$, plus the state of all the
node instances of $n$.

\subsubsection{Modeling recurrent neural networks}

We start to showcase the expressive power of \|fby| by modeling
recurrent neural networks (RNNs), taking as example the classical Long
Short-Term Memory (LSTM) \cite{lstm2}. RNNs are stateful networks,
where one cycle passes a state that is used in the next. We provide in
\figref{lstm} an implementation of the LSTM later.\footnote{Node \texttt{dense1} provides here a more faithful implementation of the Keras layer \texttt{Dense} than the node \texttt{dense} of \figref{gdescent}.}

The presentation of the LSTM layer is meant to facilitate its comparison with classical ML frameworks such as Keras, preserving the separation between the \emph{cell} node \|lstm_cell| representing the function iterated within the recurrence from the \|lstm| node which implements the reduction over time.
Parameter initialization follows the Keras default: \texttt{glorot\_uniform} for the kernel, zeros for the bias, \texttt{orthogonal} for the recurrent state.
Notice how the two LSTM-specific recurrence state elements (\|c| and \|h|) are encoded in node \|lstm| using two \|fby| equations.

\def\CCB{\CB}
\def\CCR{\CR}
\def\CCG{\CG}
\begin{figure}
\vspace*{-5mm}
\begin{lstlisting}[basicstyle=\ttfamily\footnotesize,lineskip=-2pt]  
node dense1(units,sizex,x)->(o)
  kernel = param(glorot_uniform([units,sizex]));
  bias = param(zeros([units]));
  o = matmul(kernel,x) + bias;
node lstm_cell(units,sizex,h,c,x)->(h_next,c_next)
  zr     = reshape([4,units],dense1(4 * units,
                     units + sizex,concat(x,h)));
  c_next = sigmoid(zr[0]) * c + sigmoid(zr[1]) * tanh(zr[3]);
  h_next = sigmoid(zr[2]) * tanh(c_next);
node lstm(units,sizex,x)->(h_next)          
  h_next,c_next = lstm_cell(units,sizex,h,c,x); 
  h = param(orthogonal([units])) fby h_next;
  c = param(orthogonal([units])) fby c_next;
\end{lstlisting}
\vspace*{-3mm}
\caption{LSTM layer (square brackets index functional arrays).}
\vspace*{-4mm}
\label{fig:lstm}
\end{figure}

Pairing the \|fby| operators within the top-level non-recurrent node is mainly due to the limitations of existing ML frameworks.
Using the node hierarchy of \REACT, we could restructure the LSTM cell to emphasize its internal organization into \emph{gates}, as defined in the literature \cite{lstm2}.
Collecting \|fby| operators for efficient iteration may deferred to the compiler (see Pompougnac et al.\ \cite{hipeac22}, Section~4).

\subsubsection{Pipelined inference}
\label{sec:pipelined}

Given the complexity of mapping ML applications onto heterogeneous hardware,
efficient designs may involve source-level mapping and scheduling \cite{huang2019gpipe,whitlock-tf-semantics}. This is cumbersome at best with existing ML frameworks.
\figref{gdescent} shows how to split node \|app| in two pipeline stages by inserting a \|fby| operator between the two layers. The result is shown in node \|app_pipe|, and the corresponding backpropagation code in node \|diff_app_pipe|.

\subsubsection{Data preprocessing}

Finally, using only function application and \|fby|, one may implement non-trivial data pre-processing typically occurring in conjunction with ML models. 
\begin{wrapfigure}{l}{0.3\textwidth}
  \begin{center}
    \vspace*{-4mm}
\def\CCB{\CB}
\def\CCR{\CR}
\def\CCG{\CG}
\begin{lstlisting}[basicstyle=\ttfamily\footnotesize,lineskip=-2pt]
node mlp(shape,x)->(o)
  y = dense1(100,shape,x);
  z = relu(x);
  o = dense1(1,100,y);
node window(x)->(w)
  x1 = 0.0 fby x;
  x2 = 0.0 fby x1;
  x3 = 0.0 fby x2;
  w  = [x,x1,x2,x3];
node timeseries(x)->(o)
  b = window(x);
  o = mlp(4,b);
\end{lstlisting}
\vspace*{-4mm}
\caption{MLP with embedded data pre-processing.}
\vspace*{-4mm}
\label{fig:mlp}
  \end{center}
\end{wrapfigure}
We showcase this in \figref{mlp} with a time series application
where inputs are buffered into a sliding window (also known as a rotating
buffer) of size 4 before being fed to a Multi-Level Perceptron (MLP).
The sliding window node uses a delay line where \|x1| is \|x|
delayed by 1 cycle, \|x2| is further delayed by 1 cycle, and finally \|x3| is
\|x| delayed by 3 cycles. The sliding window output \|w| is
obtained by grouping all four variables into a vector.\footnote{Specific language constructs, like TensorFlow's \|stop_gradient|, are needed to avoid backpropagation through the data processing code, but these mechanisms are outside the scope of our paper.}

Beyond this small example, data pre-processing and post-processing can be arbitrarily complex, from signal processing to extract a spectral representation to using the output of the ML algorithm for model-predictive control.

\subsection{Conditional execution and sampling}
\label{sec:conditional}

Piecewise functions (e.g.\ ReLU) and data sampling
have long been part of ML practice.
Yet general primitives such as the \|tf.where| choice operator or \|tf.cond|/\|xla.conditional|/\|jax.lax.cond| for conditional execution\footnote{\url{https://jax.readthedocs.io/en/latest/_autosummary/jax.lax.cond.html}} are rarely used directly by programmers.
However, the pressure to reduce the computational requirements of ML applications naturally results in rapidly growing popularity of conditional execution, such as the sparsely-gated mixture-of-experts layer \cite{sparsely-gated}.

By comparison, reactive programming exposes powerful conditional execution primitives.
The danger associated with conditional execution in a concurrent system is non-determinism: when the producer and the consumer of a variable are executed on different conditions, the consumer may read an undefined value.
This is why dataflow reactive formalisms propose correct-by-construction semantics ensuring that all values used in computations are well-defined.
This is the approach we follow in \REACT, borrowing the \|when| and \|merge| conditional execution primitives of Lustre \cite{lustreRTSS,pouzetLustre}.

\begin{figure}
  \vspace*{-6mm}
  \begin{center}
    {\footnotesize
\begin{tikzpicture}
    [node distance=1cm]     
\matrix (first) [table]
{
cycle & 0                  & 1                    & 2                   & 3             & 4                  \\
c     & \node(y0){\T}; &  \node(y1){\F}; &  \node(y2){\T}; &  \node(y3){\CG$\ABS$}; & \node(y4){\F}; \\
y     & \node(y0){$2$};    &  \node(y1){$7$};     &  \node(y2){$5$};    &  \node(y3){\CG$\ABS$}; & \node(y4){$1$}; \\
x     & \node(y0){$2$};    &  \node(y1){\CG$\ABS$};        &  \node(y2){$5$};    &  \node(y3){\CG$\ABS$}; & \node(y4){\CG$\ABS$}; \\
};
\draw [->,brown,thick] (-1.4,        0.1) -- (-1.4,       -0.7);
\draw [->,brown,thick] (-1.4+2*1.05, 0.1) -- (-1.4+2*1.05,-0.7);
\draw [->,brown,thick] (-2.0,       -0.3) -- (-2.0,       -0.7);
\draw [->,brown,thick] (-2.0+2*1.05,-0.3) -- (-2.0+2*1.05,-0.7);
\end{tikzpicture}}
        \vspace*{-3mm}
    \caption{Sample chronogram of \|x = y when c|.}
        \vspace*{-4mm}
    \label{fig:chronowhen}
\end{center}
\end{figure}

In the dataflow setting of Lustre and \REACT, where execution is driven by the arrival of data, conditioning the execution of a node in a cycle is done by making its input variables present or absent in that cycle. This is done using the (sub-)sampling operator \|when|. In equation ``\|x = y when c|'', it is required that variable \|c| is of Boolean type and that \|c| is present in a cycle if and only if \|y| is present. 
Then, in cycles where \|c| is present with value $\T$, \|x| takes the value of \|y|.
In cycles where \|c| is absent or present with value $\F$, \|x| is absent and computations that depend on it are not executed. If \|c| is absent, then \|x| is absent. We provide in \figref{chronowhen} a possible execution trace. In cycles 0 and 2, \|c| is present with value $\T$, so the value of
\|y| is copied onto \|x|. In cycles~1 and~4, both inputs are present but \|c| is $\F$ so \|x| is absent. Finally, all variables are absent in cycle~3.

Primitive ``\|x = merge c y z|'' defines a deterministic multiplexer. Its inputs must satisfy the following conditions:
\begin{itemize}
\item \|c| is known as the \emph{control stream} of the merge operator and must be of Boolean type;
\item \|y| is be present in a cycle \emph{iff} \|c| is present with value $\T$;
\item \|z| is be present in a cycle \emph{iff} \|c| is present with value $\F$;
\item \|y| and \|z| have the same data type.
\end{itemize}

\begin{wrapfigure}{r}{0.5\textwidth}
  \vspace*{-8mm}
  \begin{center}
    {\footnotesize
\begin{tikzpicture}
    [node distance=1cm]     
\matrix (first) [table]
{
cycle & 0                  & 1                    & 2                   & 3             & 4                   \\
c     & \node(y0){\T}; &  \node(y1){\F}; &  \node(y2){\T}; &  \node(y3){\CG$\ABS$}; & \node(y4){\F}; \\
y     & \node(y0){$2$};    &  \node(y1){\CG$\ABS$};        &  \node(y2){$5$};    &  \node(y3){\CG$\ABS$}; & \node(y4){\CG$\ABS$};        \\
z     & \node(y0){\CG$\ABS$};       &  \node(y1){$3$};     &  \node(y2){\CG$\ABS$};       &  \node(y3){\CG$\ABS$}; & \node(y4){$1$};     \\
x     & \node(y0){$2$};    &  \node(y1){$3$};     &  \node(y2){$5$};    &  \node(y3){\CG$\ABS$}; & \node(y4){$1$};     \\
};
\draw [->,brown,thick] (-1.4,        0.4) -- (-1.4,       -0.8);
\draw [->,brown,thick] (-1.4+1*1.05, 0.4) -- (-1.4+1*1.05,-0.8);
\draw [->,brown,thick] (-1.4+2*1.05, 0.4) -- (-1.4+2*1.05,-0.8);
\draw [->,brown,thick] (-1.4+4*1.05, 0.4) -- (-1.4+4*1.05,-0.8);
\draw [->,brown,thick] (-2.0,       -0.1) -- (-2.0,       -0.8);
\draw [->,brown,thick] (-2.0+2*1.05,-0.1) -- (-2.0+2*1.05,-0.8);
\draw [->,brown,thick] (-2.0+1*1.05,-0.5) -- (-2.0+1*1.05,-0.8);
\draw [->,brown,thick] (-2.0+4*1.05,-0.5) -- (-2.0+4*1.05,-0.8);
\end{tikzpicture}}
        \vspace*{-3mm}
    \caption{Sample chronogram of \|x = merge c y z|.}
        \vspace*{-4mm}
    \label{fig:chronomerge}
\end{center}
\end{wrapfigure}

Under these conditions, \|x| is a variable that is present \emph{iff} \|c| is present, which takes the value of \|y| if \|c| is $\T$ and the value of \|z| if \|c| is $\F$.
\figref{chronomerge} shows a possible execution trace. In cycles 0 and 2, \|c| is present with value $\T$, so the value of
\|y| is copied onto \|x| and \|z| must be absent. In cycles 1 and 4, \|c| is present with value $\F$, so the value of
\|z| is copied onto \|x| and \|y| must be absent. In cycle 3, all variables are absent.

As a first ML application of \|when| and \|merge|,
\figref{moe} describes a sparsely-gated mixture-of-experts layer \cite{sparsely-gated}.
We instantiate it for 3 experts, assuming that they are provided as nodes \|expert1|, \|expert2|, and \|expert3|, which produce tensors of the same shape.
\begin{wrapfigure}{l}{0.45\textwidth}
        \vspace*{-4mm}
\begin{lstlisting}[basicstyle=\ttfamily\footnotesize,lineskip=-2pt]
node gating_network(x)->(g1,g2,g3)
node expert1(x)->(o)
node expert2(x)->(o)
node expert3(x)->(o)

node mixture_of_experts(shape,x)->(o)
  g1,g2,g3 = gating_network(x);
  o1 = expert1(x when c1); c1 = (g1 != 0); 
  o2 = expert2(x when c2); c2 = (g2 != 0); 
  o3 = expert3(x when c3); c3 = (g3 != 0);
  o  = g1 * (merge c1 o1 zeros(shape)) +
       g2 * (merge c2 o2 zeros(shape)) +
       g3 * (merge c3 o3 zeros(shape));
\end{lstlisting}
        \vspace*{-4mm}
  \caption{Sparsely-gated Mixture of Experts layer.}
        \vspace*{-4mm}
  \label{fig:moe}
\end{wrapfigure}
We assume that expert control is performed by node \|gating_network|, which outputs at each cycle one floating point value per expert, serving as weights in the computation of the output (in lines 11-13).
In a cycle where the weight corresponding to an expert is 0, the expert must not be computed. This behavior is obtained by sampling the network input \|x|.
For \|expert1| this is done in line 8. Boolean variable \|c1| stores the Boolean activation condition.
To compute the weighted sum output, \|o1|, \|o2|, and \|o3| cannot be added because they are not present at the same cycle.
Instead, the \|merge| operators of lines 11-13 extend the outputs \|o1|, \|o2|, and \|o3| to variables that are present at all cycles, padding with 0 in cycles where the experts do not produce a value.

Notice that Lustre and \REACT offer a modular form of logical time, by allowing subsets of equations, including \|fby| operators, to be executed under specific Boolean conditions.
For instance, the experts or the gating network of \figref{moe} can be themselves stateful/recurrent \cite{RMoE} involving \|fby| operators.
For every such expert, the recurrences relate only the cycles where the expert is active.
In the context of control applications---typically multi-periodic and involving multiple execution modes---this facilitates interfacing the ML code with the reactive system-level code driving it \cite{multirate,multirate2}.

\subsection{Backpropagation training of feedforward networks}
\label{sec:advanced}
\label{sec:backprop}

We are now ready to discuss the reactive modeling of gradient backpropagation when training neural networks.
In the example of \figref{gdescent}, node \|diff_app| implements
backpropagation for the feedforward network \|app|. Note that
\|diff_app| preserves all the inputs, outputs, and equations of
\|app|, structurally adding to them the error computation and backward
path variables and equations and the control governing them.  The new
Boolean input \|bp|, known as the \emph{backpropagation signal}, is
present at all cycles. Backpropagation is performed in cycles where
\|bp| is $\T$. In these cycles, the node also receives a ground truth
value \|gt|.
Based on the value of \|gt|, the error is computed (on the current sample
only), it is multiplied by the learning rate, and then backpropagated.
Parameters are eventually updated, with the objective of reducing
the error in future cycles. 

Note that an execution of node \|diff_app| can mix cycles of
training with cycles of pure inference, by choosing the sequence of values of
\|bp|. This feature is needed in
Reinforcement Learning \cite{Dul21}, and our modeling approach
supports it out-of-the-box. Pure inference and pure training
implementations of our specification (like those produced by
TensorFlow) can then be obtained by setting \|bp| to constant
$\F$ or to constant $\T$, respectively, as we do in nodes
\|infer| and \|train|.

The synthesis of the backpropagation and parameter update code from
the specification is not detailed in this paper.\footnote{And neither are, for of the sake of the presentation presentation, the issues of selectively saving or loading the state of \texttt{fby} primitives.}
We only provide the result of this synthesis process to show that \REACT allows its representation.
Backpropagation consists in computing, for each value $v$ computed on
the forward path, an error estimation \|d$v$|. In our backpropagation
approach based on static program transformations, for each
specification-level differentiable variable $v$ we create in the
backpropagating implementation another variable \|d$v$|.  The
computation of the \|d$v$| variables reverses all the causalities of the
forward path. For instance, in node \|dense| of \figref{gdescent}
variable \|o| is computed from \|k|, \|i|, and
\|b|. Then, in node \|diff_dense|, variables \|dk|,
\|di|, and \|db| all depend on \|do|. Furthermore,
backward path computations may depend on forward path values (when the
forward path function is not affine). For instance, \|dk| depends
on \|i|.
The forward path depends on the backward path through the update of parameters only.
This explains the form of the \|diff_param| node, which accumulates the corrections \|do| at every cycle where \|bp| is present.
The updated parameter value is fed to the next cycle regardless of the value of \|bp|.

From a language design perspective, \|param| is an ML-specific
primitive (the only one used for the scope of this paper), which is
expanded during synthesis of training code (not detailed in this
paper) into \|diff_param|. Further coverage of ML parameters, from
a recurrent reactive specification perspective, follows in
\secref{bptt}.

\section{Recurrences backward in time}
\label{sec:revtime}

The reactive core of \REACT, along with the time/space and state manipulation extensions of the previous section, already covers the expressiveness requirements of ML applications in the yellow cells of \figref{complexityclasses}.
The particularity of these ML applications is that they only define data dependences (recurrences) forward in time---a value of a cycle cannot depend on values of future cycles.

\begin{figure}
  \vspace*{-5mm}
  \begin{center}
    {\footnotesize
\begin{tikzpicture}
    [node distance=1cm]     
\matrix (first) [table]
{
cycle & 0                 & 1                  & 2                 & 3                  & 4                & 5  \\
x     & \node(z0){$4.3$}; &  \node(z1){$3.0$}; &  \node(z2){\CG$\ABS$}; &  \node(z3){$3.3$}; & \node(z4){$1.9$}; & \node(z5){$7.7$};    \\
y     & \node(x0){$3.0$}; &  \node(x1){$3.3$}; &  \node(x2){\CG$\ABS$}; &  \node(x3){$1.9$}; & \node(x4){$7.7$}; & \node(x5){$1.9$};   \\
};
\draw [->,red,thick] (-1.4,        0.0) -- (-1.9,       -0.4);
\draw [->,red,thick] (-1.4+2*1.05, 0.0) -- (-1.9+  1.05,-0.4);
\draw [->,red,thick] (-1.4+3*1.05, 0.0) -- (-1.9+3*1.05,-0.4);
\draw [->,red,thick] (-1.4+4*1.05, 0.0) -- (-1.9+4*1.05,-0.4);
\draw [->,red,thick] (-1.4+5*1.05, 0.0) -- (-1.9+5*1.05,-0.4);
\end{tikzpicture}}
        \vspace*{-3mm}
   \caption{Sample chronogram of \|x = post y|. In red, data dependences.}
        \vspace*{-4mm}
    \label{fig:chronopost}
\end{center}
\end{figure}

But to cover the needs of modern Machine Learning, we also need to cover the applications in the orange cells of \figref{complexityclasses}.
These applications feature dependences backwards in time coming from 3 main sources: bidirectional networks \cite{brnn}, batch normalization \cite{batchNorm}, and backpropagation through time used in the training and reinforcement learning of recurrent networks \cite{fstbptt,bptt}.

To this end, we extend reactive dataflow programming with a new primitive, with the form ``\verb|y = post x;|''.
This equation specifies that \|x| and \|y| are present at the same cycles
and that, if we denote with $t_0<t_1<t_2<\ldots$ the indices of the cycles where \|x| and \|y| are both present, then for all
$n$ we have $y_{t_n}=x_{t_{n+1}}$.
\figref{chronopost} shows one possible execution trace for this equation.
We start our exploration of the new primitive by noticing that ``\|post (y fby x)|'' is always equal to \|x|, whereas ``\|y fby (post x)|'' is equal to \|x| for all cycles save the first, where it is equal to \|y| (since \|post| discards the first value of \|x| which cannot be recovered).

\subsection{Backpropagation on pipelined applications}

The simplest use of \|post| is in extending the backpropagation
method of \secref{backprop} to pipelined applications. Recall from
\secref{pipelined} that \|fby| primitives can be introduced in
feedforward specifications to mark the frontier between pipeline
stages of the inference algorithm.

These \|fby| primitives do not create cycles in the dataflow of
the node, and backpropagation code can be synthesized for such
specifications without considering the full complexity of time
recurrences, which is covered in the next section. More precisely,
backpropagation in pipelined feedforward applications simply requires
the realignment of backward propagation with the forward one by means
of \|post| primitives inserted on the backwards path in places
where on the forward path a \|fby| primitive exists.

\begin{wrapfigure}{r}{0.55\textwidth}
  \vspace*{-10mm}
  \begin{center}
{\footnotesize
\begin{tikzpicture}
    [node distance=1cm]     
\matrix (first) [table]
{
cycle & 0                 & 1                  & 2                 & 3                  & 4                & 5  \\
i     & \node(z0){$i_0$}; &  \node(z1){$i_1$}; &  \node(z2){$i_2$}; &  \node(z3){\CG$\ABS$};      & \node(z4){$i_4$};& \node(z5){$i_5$};    \\
o     & \node(x0){0.0};   &  \node(x1){$i_0$}; &  \node(x2){$i_1$}; &  \node(x3){\CG$\ABS$};      & \node(x4){$i_2$}; & \node(x5){$i_4$};   \\
bpo   & \node(x0){\F}; &  \node(x1){\F}; &  \node(x2){\T}; &  \node(x3){\CG$\ABS$};      & \node(x4){\F}; & \node(x5){\T};   \\
do    & \node(x0){\CG$\ABS$};      &  \node(x1){\CG$\ABS$};      &  \node(x2){$do_2$};&  \node(x3){\CG$\ABS$};     & \node(x4){\CG$\ABS$};      & \node(x5){$do_5$};   \\
bpi   & \node(x0){\F}; &  \node(x1){\T}; &  \node(x2){\F}; &  \node(x3){\CG$\ABS$};      & \node(x4){\T};  & \node(x5){\F};   \\
di    & \node(x0){\CG$\ABS$};      &  \node(x1){$do_2$}; &  \node(x2){\CG$\ABS$};     &  \node(x3){\CG$\ABS$};      & \node(x4){$do_5$}; & \node(x5){\CG$\ABS$};   \\
};
\draw [->,blue,thick] (-2.0,        0.8) -- (-1.5,        0.5);
\draw [->,blue,thick] (-2.0+1*1.05, 0.8) -- (-1.5+1*1.05, 0.5);
\draw [->,blue,thick] (-2.0+2*1.05, 0.8) -- (-1.5+3*1.05, 0.5);
\draw [->,blue,thick] (-2.0+4*1.05, 0.8) -- (-1.5+4*1.05, 0.5);
\draw [->,red,thick] (-1.5+0*1.05, 0.0) -- (-2.0+0*1.05,-0.8);
\draw [->,red,thick] (-1.5+1*1.05, 0.0) -- (-2.0+1*1.05,-0.8);
\draw [->,red,thick] (-1.5+3*1.05, 0.0) -- (-2.0+2*1.05,-0.8);
\draw [->,red,thick] (-1.5+4*1.05, 0.0) -- (-2.0+4*1.05,-0.8);
\draw [->,red,thick] (-1.5+5*1.05, 0.0) -- (-2.0+5*1.05,-0.8);
\draw [->,red,thick] (-1.5+1*1.05,-0.4) -- (-2.0+1*1.05,-1.3);
\draw [->,red,thick] (-1.5+4*1.05,-0.4) -- (-2.0+4*1.05,-1.3);
\end{tikzpicture}}
  \vspace*{-4mm}
    \caption{Backpropagation in pipelined applications: a possible trace of node
      \texttt{diff\_fby} of \figref{gdescent}. In blue, dependences
      forward in time. In red, dependences backward in
      time.}
    \label{fig:bppipe}
  \end{center}
  \vspace*{-4mm}
\end{wrapfigure}

We illustrate this on \figref{gdescent}: \|app_pipe| is a pipelined version of \|app|, and
\newline\noindent \|diff_app_pipe| shows the effect of injecting backpropagation code.
In the process, the unique \|fby| primitive is replaced with an instance
of the \|diff_fby1| node which not only includes the
forward-path \|fby| that separates the pipeline stages,
but also includes the backward-path \|post| primitives
that realign backpropagated values with the forward path computations.\footnote{To simplify the code in \figref{gdescent}, we assume in the definition of \texttt{diff\_fby1} that its first argument is a constant, requiring no backpropagation. A full-fledged version of \texttt{fby} backpropagation is provided in \secref{bptt}.}

In this chronogram, the value $i_1$ is produced by the first pipeline
stage (node \|dense| in our example) in cycle 1, but is delayed to
be processed by the second pipeline stage in cycle 2. In cycle 2, the
backpropagation indicator is $\T$ in the second stage
($\|bpo|=\T$) and value $do_2$ must be backpropagated to the
cycle 1 of the first pipeline stage. This is realized by the two
\|post| primitives in node \|diff_fby1|. The first one
simply anticipates the values of \|bpo| to produce
\|bpi|. The second is part of a more complex protocol that
produces an output \|di| at a different rate from the
input \|do|.

\subsection{Globally-forward, locally-backward propagation}
\label{sec:gflb}

Note that, while the \|fby| and \|post|
primitives of \REACT both embody recurrences over time, they are not
symmetrical, as classical functional programming iterators
\|fold_left| and \|fold_right| are.  As any execution has a
start (in cycle 0) forward recurrences always have a well-defined
initialization cycle.  By comparison,
reactive programming does not identify an end cycle and executions
are \emph{a priori} infinite.  For this reason, no initial value is
specified by the \|post| primitive.

Like for \|fby|, the output of a \|post| does not depend on
its input inside a cycle. This allows the creation of cyclic dataflow graphs, like
the \|counter| node of \figref{counter}, which represent
recurrences in time. Replacing in the
\|counter| node the \|fby| by a \|post| produces the node
of \figref{future}(left). We have pictured in \figref{future}(right)
the data dependences specified by its equations.\footnote{Brown
dependences inside a cycle are due to the \texttt{+} operator.
Red dependences between cycles are associated with \texttt{post}.}
Clearly, the program can compute no output, as all values depend on
the infinite future.

Whenever a program includes a \|post|
primitive in a dataflow cycle, we must ensure that, at every
execution cycle in the program execution, the dependence on
the future is eventually broken.
In formal terms, the predicate $p$ identifying the cycles
where the dependence on the future is broken must satisfy the
temporal logic \cite{tl} formula \textit{always eventually $p$}.

\begin{figure}[h!tb]
  \begin{center}
    \begin{tabular}{ccc}
      \begin{minipage}{4cm}
\def\CCB{\CB}
\def\CCR{\CR}
\def\CCG{\CG}
\begin{lstlisting}[basicstyle=\ttfamily\footnotesize]
  node pcounter()->(o)
    o = post u;
    u = o + 1;
\end{lstlisting}
      \end{minipage} & &
      \begin{minipage}{8cm}
{\footnotesize
\begin{tikzpicture}
    [node distance=1cm]     
\matrix (first) [table]
{
cycle & 0                 & 1                  & 2                 & 3                  & 4                & 5  \\
u     & \node(z0){$u_0$}; &  \node(z1){$u_1$}; &  \node(z2){$u_2$}; &  \node(z3){$u_3$}; & \node(z4){$u_4$}; & \node(z5){$u_5$};    \\
o     & \node(x0){$o_0$}; &  \node(x1){$o_1$}; &  \node(x2){$o_2$}; &  \node(x3){$o_3$}; & \node(x4){$o_4$}; & \node(x5){$o_5$};   \\
};
\draw [->,brown,thick] (-2.1,       -0.4) -- (-2.1,       -0.1);
\draw [->,brown,thick] (-2.1+1*1.05,-0.4) -- (-2.1+1*1.05,-0.1);
\draw [->,brown,thick] (-2.1+2*1.05,-0.4) -- (-2.1+2*1.05,-0.1);
\draw [->,brown,thick] (-2.1+3*1.05,-0.4) -- (-2.1+3*1.05,-0.1);
\draw [->,brown,thick] (-2.1+4*1.05,-0.4) -- (-2.1+4*1.05,-0.1);
\draw [->,brown,thick] (-2.1+5*1.05,-0.4) -- (-2.1+5*1.05,-0.1);
\draw [->,red,thick] (-1.4,        0.0) -- (-1.9,       -0.5);
\draw [->,red,thick] (-1.4+1*1.05, 0.0) -- (-1.9+1*1.05,-0.5);
\draw [->,red,thick] (-1.4+2*1.05, 0.0) -- (-1.9+2*1.05,-0.5);
\draw [->,red,thick] (-1.4+3*1.05, 0.0) -- (-1.9+3*1.05,-0.5);
\draw [->,red,thick] (-1.4+4*1.05, 0.0) -- (-1.9+4*1.05,-0.5);
\draw [->,red,thick] (-1.4+5*1.05, 0.0) -- (-1.9+5*1.05,-0.5);
\end{tikzpicture}}
      \end{minipage}
    \end{tabular}
    \vspace*{-4mm}
    \caption{Dependences on the infinite future; backward
      dependences in red, instantaneous ones in brown.}
    \label{fig:future}
  \end{center}

  \begin{center}
    \begin{tabular}{ccc}
      \begin{minipage}{5cm}
\def\CCB{\CB}
\def\CCR{\CR}
\def\CCG{\CG}
\begin{lstlisting}[basicstyle=\ttfamily\footnotesize]
node backfill(i,bp)->(o)
  o = merge bp
        (i when bp)
        ((post o) when not bp);
\end{lstlisting}
      \end{minipage} & &
      \begin{minipage}{7cm}
{\footnotesize
\begin{tikzpicture}
    [node distance=1cm]     
\matrix (first) [table]
{
cycle & 0                 & 1                  & 2                 & 3                  & 4                & 5  \\
end   & \node(z0){\F}; &  \node(z1){\F}; &  \node(z2){\T}; &  \node(z3){\T}; & \node(z4){\F}; & \node(z5){\T};    \\
i     & \node(z0){2};     &  \node(z1){6};     &  \node(z2){8};    &  \node(z3){5};    & \node(z4){1};     & \node(z5){4};    \\
o     & \node(x0){8};     &  \node(x1){8};     &  \node(x2){8};    &  \node(x3){5};    & \node(x4){4};     & \node(x5){4};   \\
};
\draw [->,brown,thick] (-2.6,        0.2) -- (-2.6,       -0.6);
\draw [->,brown,thick] (-2.6+1*1.05, 0.2) -- (-2.6+1*1.05,-0.6);
\draw [->,brown,thick] (-2.6+2*1.05, 0.2) -- (-2.6+2*1.05,-0.6);
\draw [->,brown,thick] (-2.6+3*1.05, 0.2) -- (-2.6+3*1.05,-0.6);
\draw [->,brown,thick] (-2.6+4*1.05, 0.2) -- (-2.6+4*1.05,-0.6);
\draw [->,brown,thick] (-2.6+5*1.05, 0.2) -- (-2.6+5*1.05,-0.6);
\draw [->,red,thick] (-1.9+2*1.05,-0.3) -- (-1.9+2*1.05,-0.6);
\draw [->,red,thick] (-1.9+3*1.05,-0.3) -- (-1.9+3*1.05,-0.6);
\draw [->,red,thick] (-1.9+5*1.05,-0.3) -- (-1.9+5*1.05,-0.6);
\draw [->,red,thick] (-2.4+1*1.05,-0.65) -- (-2.1+0*1.05,-0.65);
\draw [->,red,thick] (-2.4+2*1.05,-0.65) -- (-2.1+1*1.05,-0.65);
\draw [->,red,thick] (-2.4+5*1.05,-0.65) -- (-2.1+4*1.05,-0.65);
\end{tikzpicture}}
      \end{minipage}
    \end{tabular}
    \vspace*{-3mm}
    \caption{Bounded broadcast towards the past.}
    \label{fig:backfill}
    \vspace*{-5mm}
  \end{center}
\end{figure}

This requirement only applies to \|post|, not to \|fby|.
Thus, the execution of the program can proceed \emph{globally forwards}, \emph{locally backwards}.

In \figref{backfill}(left), the condition breaking the dependence on the future is provided by the \|bp| input. When \|bp| is $\T$, the value
of the input is sampled at the current cycle and broadcasted to all
previous cycles, until reaching an earlier cycle where \|bp| was $\T$.
One execution trace of \|backfill| is pictured
in \figref{backfill}(right). Notice how the cycles
where \|bp| is $\T$ cut the execution into intervals, with
the output remaining constant over each interval.
\subsubsection{Batch normalization}
\label{sec:batchnorm}

We are now ready to study the \emph{batch normalization} layer \cite{DBLP:journals/corr/IoffeS15}, which is one of the most complex from a reactive control perspective.
In training mode, the forward path is provided in \figref{batchnorm} by node \|batch_norm|.

\begin{figure}
  \vspace*{-4mm}
\lstset{moredelim=[s][\color{gray}]{(*}{*)}}
\def\CCB{\CB}
\def\CCR{\CR}
\def\CCG{\CG}
\begin{lstlisting}[basicstyle=\ttfamily\footnotesize,lineskip=-2pt]
node fby_end(shape,end,init,i)->(o) 
  i1 = if end then zeros(shape) else i;
  o = if (true fby end) then init else (init fby i1);
node mean_var(shape,x,batch_end)->(mean,var)
  (* fw computation of batch size *)
  cnt     = 1 + fby_end([],batch_end,0,cnt);
  (* fw computation of batch sum *)
  sum     = x + fby_end(shape,batch_end,zeros(shape),sum); 
  (* backfill batch mean *)
  mean    = backfill(sum / cnt,batch_end);                 
  var_sum = (x - mean) * (x - mean) +
            fby_end(shape,batch_end,zeros(shape),var_sum); 
  (* backfill batch variance *)
  var     = backfill(var_sum / cnt,batch_end);             
node norm(shape,epsilon,mean,var,x)->(y)
  gamma = fby_end(shape,batch_end,param(ones(shape)),gamma);
  beta = fby_end(shape,batch_end,param(zeros(shape)),beta);
  y = gamma * (x - mean) / (var + epsilon) + beta;
node batch_norm(shape,epsilon,x,batch_end)->(y)
  mean,var = mean_var(shape,x,batch_end);
  y = norm(shape,epsilon,mean,var,x);
\end{lstlisting}
\vspace*{-3mm}
  \caption{Batch normalization.}
  \label{fig:batchnorm}
\vspace*{-3mm}
\end{figure}

We operate on batches of samples of the input \|x|.
Batch termination is identified by the Boolean input \|batch_end|.
Node \|mean_var| computes the mean and variance, propagating data both forward and backward in time.
The number of samples in a batch and the partial sum of its elements are accumulated in the forward direction in variables \|cnt| and \|sum|, respectively.
Node \|fby_end| applies forward in time recurrences piecewise:
whenever \|end| is $\T$, the input is dropped and in the next cycle the output is set to \|init|.
In other cycles, the output is set to the delayed input.
An execution trace of \|fby_end| is provided in \figref{fbyend}.

\begin{figure}
  \begin{center}
{\footnotesize
\begin{tikzpicture}
    [node distance=1cm]     
\matrix (first) [table]
{
cycle & 0                 & 1                  & 2                 & 3                  & 4                & 5  \\
end   & \node(z0){\F}; &  \node(z1){\F}; &  \node(z2){\T}; &  \node(z3){\T}; & \node(z4){\F}; & \node(z5){\T};    \\
init  & \node(z0){1};     &  \node(z1){0};     &  \node(z2){9};    &  \node(z3){3};    & \node(z4){5};     & \node(z5){8};    \\
i     & \node(z0){2};     &  \node(z1){6};     &  \node(z2){8};    &  \node(z3){5};    & \node(z4){1};     & \node(z5){4};    \\
o     & \node(x0){1};     &  \node(x1){2};     &  \node(x2){6};    &  \node(x3){3};    & \node(x4){5};     & \node(x5){1};   \\
};
\draw [->,brown,thick] (-2.6,        0.0) -- (-2.6,       -0.8);
\draw [->,brown,thick] (-2.6+3*1.05, 0.0) -- (-2.6+3*1.05,-0.8);
\draw [->,brown,thick] (-2.6+4*1.05, 0.0) -- (-2.6+4*1.05,-0.8);
\draw [->,red,thick] (-2.1+0*1.05,-0.4) -- (-2.3+1*1.05,-0.8);
\draw [->,red,thick] (-2.1+1*1.05,-0.4) -- (-2.3+2*1.05,-0.8);
\draw [->,red,thick] (-2.1+4*1.05,-0.4) -- (-2.3+5*1.05,-0.8);
\end{tikzpicture}}
\vspace*{-3mm}
    \caption{Trace of node \|fby\_end| of \figref{batchnorm}.}
    \label{fig:fbyend}
  \end{center}
\vspace*{-4mm}
\end{figure}

In node \|mean_var|, instances of \|fby_end| are used to reset the accumulation of partial sums at batch ends, so that for each batch starts with an accumulator set to 0.
At batch ends, the value of expression ``\|sum/cnt|''---the batch mean---is sampled and broadcast to all batch cycles by the first instance of \|backfill| as described above.
Using the batch mean, one may again accumulate in forward direction, in variable \|var_sum|, the partial sums needed to compute the variance.
At batch end, the variance is broadcast to the whole batch by the second instance of \|backfill|.

The computed batch mean and variance are then used to normalize the input, also using two trainable parameters \|gamma| and \|beta|.
The use of \|fby_end| in the context of trainable parameters will be covered in detail in the next section.

\subsection{Backpropagation through time for RNN training}
\label{sec:bptt}
Consider the \|lstm| node of \figref{lstm}, which represents the
recurrent layer LSTM.  In its first execution cycle, the state
accumulator variables \|h| and \|c| are initialized with the
value of the two parameters. Then, at every cycle (including the
first), the non-recurrent \|lstm_cell| node is executed once,
taking as input the current accumulator and the current value of the
\|lstm| input \|x|, and producing the updated accumulators (of
which one is also output). If we denote with $h_n$ the value of
\|h| at cycle $n$, the computation of $h_{n+1}$ causally depends
on that of $h_n$ for all $n$.

\begin{figure}
  \vspace*{-4mm}
\begin{lstlisting}[basicstyle=\ttfamily\footnotesize,lineskip=-2pt]
node param_end(shape,end,p)->(o)
  o = o1; o2 = o1;
  o1 = fby_end(shape,end,p,o2);
node dense1_end(units,sizex,x,end)->(o)
  shape = [units,sizex]; 
  kernel = param_end(shape,end,
               param(glorot_uniform(shape)));
  bias = param_end([units],end,
               param(zeros([units])));
  o = matmul(kernel,x) + bias;
node lstm_cell_end(units,sizex,h,c,x,end)->(h_next,c_next)
  zr     = reshape([4,units],dense1_end(4 * units,
                     sizex + units,concat(x,h),end));
  c_next = sigmoid(zr[0]) * c + sigmoid(zr[1] * tanh(zr[3]));
  h_next = sigmoid(zr[2]) * tanh(c_next);
node lstm_end(units,sizex,x,end)->(h_next)          
  h_next,c_next = lstm_cell_end(units,sizex,h,c,x,end); 
  h = fby_end(end,param(orthogonal([units])),h_next);
  c = fby_end(end,param(orthogonal([units])),c_next);
\end{lstlisting}
  \vspace*{-4mm}
\caption{LSTM of \figref{lstm} extended with backpropagation reset.}
  \vspace*{-4mm}
\label{fig:lstmbackprop}
\end{figure}

Recall that in a backpropagation training context, for every value $v$ computed on the forward path, an error estimation $\textit{dv}$ is computed on the backward path, and the computation of the latter reverses all the causalities of the forward path.
Thus, if the difference associated with $h_n$ is denoted $\textit{dh}_n$, we know that $\textit{dh}_n$ depends on $\textit{dh}_{n+1}$ for all $n$, and so the computation of $\textit{dh}_0$ (to update parameters) depends on the infinite future.

As explained in \secref{gflb}, such a dependence on the infinite future cannot be effectively resolved: backpropagation must be reduced to a globally forward,
locally-backward propagation, where the predicate determining when to restart
backpropagation is \emph{always eventually} $\T$.
Such a predicate is readily available when training with mini-batches.
Indeed, RNN training samples consist of finite input sequences accompanied by ground truth values.
In a training or RL, sequences are provided cylically, possibly separated by samples meant for inference only .
The result can be seen as a single stream of inputs, where the end of training sequences is marked by the Boolean input signal \|end|.

\begin{wrapfigure}{l}{0.6\textwidth}
  \vspace*{-4mm}
\def\CCB{\CB}
\def\CCR{\CR}
\def\CCG{\CG}
\begin{lstlisting}[basicstyle=\ttfamily\footnotesize,lineskip=-2pt]
node diff_fby_end(shape,end,init,i,bp,do)->(o,dinit,di)
  o = fby_end(end,init,i);
  rst1 = rst when bp; end1 = end when bp; 
  dinit = merge rst1 (do when rst1) zeros(shape);
  di = merge end1 zeros(shape) (di1 when not end1);
  di1 = post (merge rst1 zeros(shape)
                         (do when not rst1));                                   
node diff_param_end(shape,end,p,bp,do)->(o,dp)
  o = o1; o2 = o1; (* dup *)
  do1 = do + do2;  (* backpropagation for dup *)
  o1,_,do2 = diff_fby_end(shape,end,p,o2,bp,do1);
node diff_lstm_end(units,sizex,x,end,bp,dh_next)->(h_next,dx)          
  h_next,c_next,dh,dc,dx = lstm_cell_end(units,sizex,h,
                            c,x,end,bp,dh_next,dc_next);
  hp = diff_param(orthogonal([units]),bp,dhp);
  cp = diff_param(orthogonal([units]),bp,dcp);
  h,dhp,dh_next = diff_fby_end(end,hp,h_next,bp,dh);
  c,dcp,dc_next = diff_fby_end(end,cp,c_next,bp,dc);
\end{lstlisting}
  \vspace*{-4mm}
\caption{LSTM backpropagation training code.}
  \vspace*{-4mm}
\label{fig:auxparam}
\end{wrapfigure}

One may finally weave backpropagation with reset into the LSTM of \figref{lstm}.
The result of this source-to-source transformation is provided in
\figref{lstmbackprop}, where \|fby| and \|post| operators are replaced with the resettable versions \|fby_end| in \figsref{batchnorm}{fbyend} and \|post_end| in \figref{lstmbi}, respectively.
Parameters also need to be wrapped into separate \|param_end| nodes, which are particular instanciations of \|fby_end| with the output fed back into their
delayed input.

The \|param_end| wrappers are required during synthesis of
backpropagation code.  Indeed, it is expected that RNN parameters are only updated once per training sequence.
During backpropagation over a sequence, parameter updates arriving at each cycle must be accumulated before applying the sum when backpropagation reaches
the first cycle of the sequence.
This is the exact behavior of the backpropagation code for node \texttt{param\_end} in \figref{auxparam}.
Notice in line 12 the accumulation of differences.
The behavior of \|diff_fby_end| ensures that the accumulated update is only applied to parameters on sequence start cycles.
The actual parameter update is performed by \|diff_param| (see \figref{gdescent}).
With these definitions, the backpropagation code of the LSTM may be obtained by structural translation, as pictured in \figref{auxparam}.

\subsection{Bidirectional RNNs}
\label{sec:bidir}
We have seen that the 5 reactive primitives of \REACT---the 4 classical ones and \|post|---allow to model batch normalization and backpropagation with reset, as well as the alternation of training and inference
within RL algorithms.
We complete our presentation of the expressive power of \REACT with the encoding of a bidirectional recurrent network \cite{lstmbidir}.

\begin{figure}
  \vspace*{-4mm}
\begin{lstlisting}[basicstyle=\ttfamily\footnotesize,lineskip=-2pt]
node post_end(shape,end,init,i)->(o)
  i1 = if (true fby end) then zeros(shape) else i;
  o = if end then init else (post i1);

node lstm_bi(units,shape,x,end)->(o)          
  h_next = lstm_end(units,shape,x,end);
  h_pred,c_pred = lstm_cell_end(units,shape,h,c,x,end); 
  h = post_end(shape,end,param(orthogonal()),h_pred);
  c = post_end(shape,end,param(orthogonal()),c_pred);
  o = h_next + h_pred;
\end{lstlisting}
\vspace*{-4mm}
\caption{Bidirectional LSTM layer.}
\label{fig:lstmbi}
\vspace*{-4mm}
\end{figure}

Bidirectional networks feature recurrences in time in both directions.
Thus, \|post| operators are \emph{needed in the model itself}.
For such a specification to be productive, it must include the
conditions allowing a globally forward, locally backward execution, as
discussed in previous sections. We assume that the Boolean input variable
\|end| of the node identifies the cycles where backward recurrences begin.

The bidirectional LSTM layer is described in
\figref{lstmbi}. It builds upon the extension of the LSTM
with backpropagation reset capability (nodes \|lstm_end|
and \|lstm_cell_end| of \figref{lstmbackprop}).
Between every two cycles where \|end| is $\T$, the layer
implements two independent recurrences in time, one forward
(implemented by the \|lstm_end| node instance) and
one backward. The outputs of the two recurrences are summed
at each cycle to produce the output of the layer.

\section{Semantics}
\label{sec:semantics}

As illustrated in the earlier examples, \REACT significantly extends the expressiveness of reactive (synchronous) languages such as Lustre.
In particular, the execution of a program involving \|post| (recurrences backwards in time) may not compute all the stream variables at a cycle $n$ before moving on to the next cycle $n+1$.

\paragraph{Comparison with Kahn networks}
As a first attempt at formalizing the semantics of \REACT, it is tempting to build upon the dataflow framework of Kahn \cite{Kahn74}: concurrent programs whose asynchronous semantics can be defined by means of compositions of \emph{monotonous stream functions}.
Unfortunately this is not so simple.
Consider, for instance, in \figref{backfill} the node \|backfill| and its execution trace.
Here, the values of \|o| are not computed as a stream, where new values are always added at the end: the value of \|o| at cycle~2 is known before those of cycles~1 and then~0.

\paragraph{Notations and conventions}
If $\DATATYPE$ is a set of values, $\DATATYPE^\ast$ denotes the set of \emph{finite streams of elements of $T$}, $\DATATYPE^\omega$ is the set of \emph{infinite streams}, and $\DATATYPE^\infty = \DATATYPE^\ast + \DATATYPE^\omega$.
To ease the presentation, we will consider a single set $\DATATYPE$ encompassing all values occurring in \REACT examples.

The empty stream is noted $\varepsilon$ and $\CONS$ is the stream concatenation operator; lifted to infinite streams inductively and allowing the contatenation of a finite stream (left-hand side operand) with a finite or infinite stream (right-hand side operand).
The length of a stream $s$ is noted $\LEN{s}$---it is a natural number or $\infty$.
The $n$-th value of stream $s$ is noted $\STH{s}{n}$, the sub-stream starting from index $n$ and ending immediately before index $m$ (i.e., $m$ excluded) is noted $\STH{s}{n:m}=s_{\{i\in\mathbb{N} \mid n\le i<m\}}$, and the suffix starting from index $n$ is noted $\STH{s}{n:}=s_{\{i\in\mathbb{N} \mid i\ge n\}}$.
For example, $\SHD{s} \CONS \STL{s}$ denotes the stream whose head is $\SHD{s}$ and tail is $\STL{s}$.

Considering a program $P$ in the syntax of \figref{syntax}, we assume $P$ has been normalized to an administrative form where every equation of consists of a single primitive.

$\VENV$ denotes an environment of stream variables.
In such an environment, $x \BIND v$ denotes the binding of variable $x$ to the value $v$, and $\VENV(x)$ denotes the value of variable $x$ in the environment $\VENV$.
$\SSEM e \ENDSEM$ is a function from environments to environments:
$\SSEM e \ENDSEM (\VENV)$ denotes the interpretation of expression $e$ applied to the environment $\VENV$.
Let $\FV(e)$ be the set of free variables of an expression $e$ and $\FV(x = e) = \FV(e) \setminus \{x\}$.
Let $\DV(x = e) = \{x\}$ denote the (left-hand side) variables bound by a an equation $\textit{eq}$.
These notations are lifted to tuples of equations, defining  $\FV(x^1,\ldots,x^d = e^1,\ldots,e^d) = \FV(e^1,\ldots,e^d) \setminus \{x^1,\ldots,x^d\}$ and $\DV(x^1,\ldots,x^d = e^1,\ldots,e^d) = \{x^1,\ldots,x^d\}$.
For the sake of conciseness, when clear from the context, we overload $x$ to denote a tuple of variables, $e$ a tuple of expressions, and $x = e$ a tuple of equations.

Variables and semantic interpretations are defined within a directed-complete partial order (dCPO or simply CPO).
We will later (somewhat loosely) refer to any CPO as a \emph{domain} in the following.
Given two CPOs $D$ and $D'$, their product $D \times D'$ and the set of Scott-continuous functions $D \to D'$ are also CPOs by considering the pointwise order.
If $\MAP$ is a continuous function from $D$ to $D$ where $\bot$ is the infimum (a.k.a.\ greatest lower bound) of $D$, we shall write $\FIX(\MAP) = \lim_{n\to\infty} \MAP^n(\bot)$ for the smallest fix point of $\MAP$ by the Kleene fixed point theorem.

\subsection{Denotational synchronous semantics}
\label{sec:sync}

We are interested in soundness guarantees about the termination of a single reaction, about the overall productivity (deadlock freedom), and on bounded memory execution.
These motivations call for a semantics suitable to offer such guarantees ``by construction'' or ``by design'' as opposed to relying on best-effort static or dynamic analyses after the fact.
This is the purpose of our denotational synchronous semantics.
Note that these guarantees are typically associated with the ability to compile the program to a statically scheduled form with in-place updates \cite{lustreRTSS}, but such concerns are outside the scope of this paper.

\paragraph{Domain of sparsely indexed streams}
One may relax the prefix ordering of the Kahn domain \cite{SKN} with explicitly indexed streams, making all streams effectively infinite and randomly indexable.
Given a natural number, one may reason about the value of an infinite stream at this specific index.
This value may be an element of $\DATATYPE$, or it may be the special \emph{unknown} value $\UNK$ capturing the fact that the actual value in $\DATATYPE$ (if any) is yet to be determined in the process of a fixed point computation.
Note that finite streams can be extended into infinite ones by concatenating $\UNK^\omega$; one may thus model finite computations while considering infinite streams only.
We also introduce an explicit \emph{absent} signal $\ABS$ and an \emph{error} value $\ERR$, and define $\TUNKABSERR = \DATATYPE \cup \{\UNK,\ABS,\ERR\}$.
$\ABS$ effectively sparsifies streams and computations, modeling the absence of a value in $\DATATYPE$ at a given index, while $\ERR$ is associated with incorrectly synchronized programs.\footnote{The notation is meant as a symbol for stream misalignment.}

One may define a simple ordering relation $\LTES$ over $\TUNKABSERR^\omega$ by lifting the flat CPO on individal values $\forall x \in \TUNKABSERR, \UNK \leq_S x\land x \leq_s \ERR$ (a.k.a.\ the Scott domain) to infinite streams.

This lets us define the \emph{sparsely indexed stream domain} $\DSYNC = (\TUNKABSERR^\omega, \LTES, \UNK^{\!\omega})$, a CPO whose join operator is denoted by $\CUPS$, with infimum $\UNK^\omega$ and supremum $\ERR^\omega$.

\paragraph{Synchronous semantics on sparsely indexed streams}
\figref{sync_semantics} provides the denotations for the synchronous semantics.
$c$ is a constant and $x$ is a stream variable;
$s$, $b$, $t$ and $f$ are streams in $\DSYNC$;
$i$ and $e$ are \REACT expressions;
$n$ is a natural integer.
The mark $\SF{}$ distinguishes the syntactic construct from its interpretation as a stream transformer.
Scalar operations are lifted to streams of values of the same type; this includes the usual unary and binary operations \|-|, \|+|, \|or|, etc.
The interpretation of a \|node| overloads the environment $\VENV$ to hold the biding a node name $f$ its implementation; conversely, the node instantiation $f(e)$ looks up the environment $\VENV$ to apply the corresponding function to the interpretation of $e$.
The projection $\pi_o(\!\textit{eq})$ retains the tuple of streams associated with the set of variables $o$ from the tuple of equations $\textit{eq}$.
The fixed point operator $\FIX$ is applied to every system of equations in a node: considering a tuple of equations $x = e$, it is applied to the function $\MAP = \lambda d. \SSEM e \ENDSEM(\VENV, x \BIND d)$ mapping the tuple of expressions $e$ to its tuple of bindings.
The environment $\VENV$ provides a value for all free variables in $e$ except for recursive definitions depending on $x$; the latter are provided with the tuple $d$ of arguments of the $\lambda$-term.
Fixed point iterations start with values in $\VENV$ already defined while $x$ takes the successive iterates $\MAP^n(\bot)$; we establish the existence and unicity of a least fixed point $\FIX(\MAP)$ by proving the Scott continuity of $\MAP$.\footnote{As a corollary of continuity, the iterates $\MAP^n(\bot)$ form an ascending chain, i.e.\ successive iterates of $\MAP$ propagate information monotonically, structurally from sub-expressions of $e$ and from input arguments bound in $V$.}

The synchronous stream transformers are provided in \figref{sync_transformers}; $u$ and $v$ denote individual values in $\DATATYPE$.
This semantics insists on providing a cycle-by-cycle interpretation amenable to bounded-memory execution.
The subscripted semantics notations $\SSEM e \ENDSEM[n]$ and $\SSEM e \ENDSEM[n:]$ denote the scalar value of expression $e$ at cycle $n$ and the suffix stream value of expression $e$ at all cycles $n$ and above, respectively.
The absence of a value is made explicit with $\ABS$.
We lift built-in functions, unary and binary operations to $\TUNKABSERR$, yielding $\UNK$ when one of more operands are $\UNK$, and yielding $\ERR$ when one or more operands are $\ABS$ or $\ERR$.
We decompose the \|fby| operator to isolate an ``initial state'' $\SF{\|fby|}$ from the ``steady state'' $\SF{\|pre|}$ triggered by the recurrent equations in \figref{sync_transformers}.

\begin{itemize}
\item The right-hand side of equations in \figref{sync_transformers} has been aligned on the $\CONS$ operator, concatenating the value at index $n$ with the rest of the stream.
  The value at index $n$ is referred to as \emph{instantaneous}.
  It is available for all computations at index $n$ and is obtained by restricting both left- and right-and-sides of the equations to this specific index.
  This instantaneous interpretation is a direct benefit of synchrony: it will form the basis for \emph{current cycle} effects when deriving an operational semantics.
  In the conventional forward-reactive case, it also makes efficient in-place evaluation possible with bounded memory limited to the values live within a given cycle---all stream values at index $n$ as well as the ``state'' $\STH{\ES}{n-1}$ associated with \|fby| operators.
  In the bidirectional case, multi-cycle dependences arise from the evaluation of $\STH{\ES}{n+1}$, challenging any in-place evaluation strategy; this is the topic of \secref{implementation}.
\item Equations in \figref{sync_transformers} provide the full definitions for infinite streams, concatenating the former with recurrences on a stream suffix such as $\STH{\ES}{n+1:}$.
  The value $\STH{\ES}{n}$ of the stream $\ES$ at index $n$ only depends on the left hand side of $\CONS$ in the right-hand-side of the equations, while the stream suffix $\STH{\ES}{n:}$ depends on the whole right-hand-side beyond the $\CONS$ operator.
\end{itemize}

All cases not matched in \figref{sync_transformers} yield error values.
Formally, this is enforced by defining any remaining non-matched case at a given index $n$ to the $\ERR$ value at index $n$; e.g.
$$
\begin{array}{r@{\;}c@{\;}r@{\,}c@{\,}l}
  \SF[n:]{\|fby|} (\ABS, v, i, e) & \equaldef & \ERR & \CONS & \SF[n+1:]{\|fby1|} (v, e) \\
  \SF[n:]{\|merge|} (\T \CONS \STH{\EB}{n+1:}, \ABS \CONS \STH{\ET}{n+1:}, \ABS \CONS \STH{\EF}{n+1:}) & \equaldef & \ERR & \CONS & \SF[n+1:]{\|merge|} (\!\STH{\EB}{n+1:}, \STH{\ET}{n+1:}, \STH{\EF}{n+1:})
\end{array}
$$
As a corollary, $\ABS$ never occurs together with present values in pointwise/scalar computations.
In particular, the merge operator enforces one of its branches to be present while the other is absent (or both are unknown).

We may now define a notion of synchrony: a program is \emph{synchronous} if the $\ERR$ value never occurs in the least fixed point solution of the semantic equations in \figref{sync_semantics};
formally, let $\TUNKABS = \DATATYPE \cup \{\UNK, \ABS\} = \TUNKABSERR \setminus \{\ERR\}$ represent non-erroneous, i.e.\ well-synchronized values, and consider a program $P$ from $d$-tuples to $d'$-tuples of streams, i.e.\ $\SSEM P \ENDSEM : \VENV^d \to \VENV^{d'}$,
$$P \textrm{ is synchronous} \iff \forall \EI \in (\TUNKABS^\omega)^d, \SSEM P \ENDSEM (\EI) \in (\TUNKABS^\omega)^{d'}$$

This is actually a problem when operating on constants that (by definition) are always present, preventing then for occuring as arguments of the merging operator.
The happens with recursive definitions involving \|post|, which are also always present.
To avoid these issues, we enforce that every constant and \|post| expression is sampled with an ``appropriate'' \|when $b$| expression before being operated upon; here ``appropriate'' calls for inferring the proper condition $b$ to enforce synchrony, which is the purpose of Lustre's clock calculus.

Finally, we define a bounded horizon version of the synchronous semantics of a program $P$ denoted as $\SSEM P \ENDSEM[][h]$, where the recursive evaluation of $\SF{\|post|}$ stops at a given index $h$: the rules are identical except that of \figref{sync_transformers} are used at index $n = h$.
We will use this bounded horizon semantics to establish forward-progress and bounded memory execution guarantees.

\begin{figure}[h!tb]
\vspace{-3mm}
\begin{eqnarr}
\SSEM c \ENDSEM (\VENV) & \equaldef & c^\omega \\
\SSEM x \ENDSEM (\VENV) & \equaldef & \VENV(x) \\
\SSEM \|$i$ fby $e$| \ENDSEM (\VENV) & \equaldef & \SF{\|fby|} (\VENV(\SSEM i \ENDSEM (\VENV)), \VENV(\SSEM e \ENDSEM (\VENV))) \\
\SSEM \|post $e$| \ENDSEM (\VENV) & \equaldef & \SF{\|post|} (\VENV(\SSEM e \ENDSEM (\VENV))) \\
\SSEM \|$e$ when $b$| \ENDSEM (\VENV) & \equaldef & \SF{\|when|} (\VENV(\SSEM e \ENDSEM (\VENV), \SSEM b \ENDSEM (\VENV))) \\
\SSEM \|merge $b$|\ \ET\ \EF\, \ENDSEM (\VENV) & \equaldef & \SF{\|merge|} (\VENV(\SSEM b \ENDSEM (\VENV), \SSEM \ET \ENDSEM (\VENV), \SSEM \EF\, \ENDSEM (\VENV)) \\
\SSEM \|$f$($e$)| \ENDSEM (\VENV) & \equaldef & \VENV(f) (\SSEM e \ENDSEM (\VENV)) \\
\SSEM \|node $f$($i$)->($o$)|\ \textit{eq} \ENDSEM (\VENV) & \equaldef & f \BIND \lambda d. \SSEM \pi_o(\!\textit{eq}) \ENDSEM (\VENV, i \BIND d) \textrm{ where } i \subseteq \FV(\!\textit{eq}) \land o \subseteq \DV(\!\textit{eq}) \\
\SSEM \|$x$ = $e$| \ENDSEM (\VENV) & \equaldef & \SSEM e \ENDSEM (\VENV, x \BIND x_\FIX) \textrm{ where } x_\FIX = \FIX (\lambda d. \SSEM e \ENDSEM (\VENV, x \BIND d))
\end{eqnarr}
\vspace{-4mm}
\caption{Synchronous semantics.}
\label{fig:sync_semantics}

\vspace{-2mm}
$$
\begin{array}{r@{\;}c@{\;}r@{\,}c@{\,}ll}
\color{gray}n\CONS\ldots & & \color{gray}n & \color{gray}\CONS & \color{gray}\textit{n\(+\)\(1\)}\CONS\ldots \\

\noalign{\smallskip}
\SF[n:]{\|fby|} (\EI, \ES) & \equaldef & \ABS & \CONS & \SF[n+1:]{\|fby|} (\EI, \ES) & \textrm{if } \STH{\EI}{n} = \STH{\ES}{n} = \ABS \\
\SF[n:]{\|fby|} (\EI, \ES) & \equaldef & \STH{\EI}{n} & \CONS & \SF[n+1:]{\|pre|} (\ES) & \textrm{if } \STH{\EI}{n}, \STH{\ES}{n} \in \DATATYPE \cup \{\UNK\} \\
\SF[n:]{\|pre|} (\ES) & \equaldef & \ABS & \CONS & \SF[n+1:]{\|pre|} (\ES) & \textrm{if } \STH{\ES}{n-1} = \ABS \\
\SF[n:]{\|pre|} (\ES) & \equaldef & \STH{\ES}{n-1} & \CONS & \SF[n+1:]{\|pre|} (\ES) & \textrm{if } \STH{\ES}{n-1} \in \DATATYPE \cup \{\UNK\} \\

\noalign{\smallskip}
\SF[n:]{\|post|} (\ES) & \equaldef & \ABS & \CONS & \SF[n+1:]{\|post|} (\ES) & \textrm{if } \STH{\ES}{n+1} = \ABS\\
\SF[n:]{\|post|} (\ES) & \equaldef & \STH{\ES}{n+1} & \CONS & \SF[n+1:]{\|post|} (\ES) & \textrm{if } \STH{\ES}{n+1} \in \DATATYPE \cup \{\UNK\} \\

\noalign{\smallskip}
\SF[n:]{\|when|} (\ABS \CONS \STH{\ES}{n+1:}, \ABS \CONS \STH{\EB}{n+1:}) & \equaldef & \ABS & \CONS & \SF[n+1:]{\|when|} (\!\STH{\ES}{n+1:}, \STH{\EB}{n+1:}) \\
\SF[n:]{\|when|} (\STH{\ES}{n} \CONS \STH{\ES}{n+1:}, \T \CONS \STH{\EB}{n+1:}) & \equaldef & \STH{\ES}{n} & \CONS & \SF[n+1:]{\|when|} (\!\STH{\ES}{n+1:}, \STH{\EB}{n+1:}) \\
\SF[n:]{\|when|} (\STH{\ES}{n} \CONS \STH{\ES}{n+1:}, \F \CONS \STH{\EB}{n+1:}) & \equaldef & \ABS & \CONS & \SF[n+1:]{\|when|} (\!\STH{\ES}{n+1:}, \STH{\EB}{n+1:}) \\
\SF[n:]{\|when|} (\STH{\ES}{n} \CONS \STH{\ES}{n+1:}, \UNK \CONS \STH{\EB}{n+1:}) & \equaldef & \UNK & \CONS & \SF[n+1:]{\|when|} (\!\STH{\ES}{n+1:}, \STH{\EB}{n+1:}) \\

\noalign{\smallskip}
\SF[n:]{\|merge|} (\ABS \CONS \STH{\EB}{n+1:}, \ABS \CONS \STH{\ET}{n+1:}, \ABS \CONS \STH{\EF}{n+1:}) & \equaldef & \ABS & \CONS & \SF[n+1:]{\|merge|} (\!\STH{b}{n+1:}, \STH{\ET}{n+1:}, \STH{\EF}{n+1:}) \\
\SF[n:]{\|merge|} (\T \CONS \STH{\EB}{n+1:}, \SHD{\ET} \CONS \STH{\ET}{n+1:}, \ABS \CONS \STH{\EF}{n+1:}) & \equaldef & \SHD{\ET} & \CONS & \SF[n+1:]{\|merge|} (\!\STH{\EB}{n+1:}, \STH{\ET}{n+1:}, \STH{\EF}{n+1:}) \\
\SF[n:]{\|merge|} (\F \CONS \STH{\EB}{n+1:}, \ABS \CONS \STH{\ET}{n+1:}, \SHD{\EF} \CONS \STH{\EF}{n+1:}) & \equaldef & \SHD{\EF} & \CONS & \SF[n+1:]{\|merge|} (\!\STH{\EB}{n+1:}, \STH{\ET}{n+1:}, \STH{\EF}{n+1:}) \\
\SF[n:]{\|merge|} (\UNK \CONS \STH{\EB}{n+1:}, \STH{\ET}{n:}, \STH{\EF}{n:}) & \equaldef & \UNK & \CONS & \SF[n+1:]{\|merge|} (\!\STH{\EB}{n+1:}, \STH{\ET}{n+1:}, \STH{\EF}{n+1:})
\end{array}
$$
%
%
\vspace{-4mm}
\caption{Synchronous stream transformers.}
\label{fig:sync_transformers}
\vspace{-3mm}
\end{figure}

\figref{illus_sync} illustrates the synchronous semantics on the \|backfill| example.
The fixed point computation is best decomposed over two dimensions: for a given stream index $m$, rows correspond to successive iterations of the fixed point computation defining values at index $m$, while the index may be stepped to $m+1$ as soon as a fixed point is reached at index $m$.

\begin{figure}[h!tb]
  \begin{minipage}[t]{.44\columnwidth}
  \scriptsize$$
  \begin{array}{|>{\SSEM}l<{\ENDSEM[\CYCLE]}|c@{\;}c@{\;}c|}
    \hline
    \noalign{\gdef\CYCLE{0:3}}
    \|i|                    &    0 &    1 &    2 \\
    \|bp|                   &   \F &   \F &   \T \\
    \hline
    \hline
    \noalign{\gdef\CYCLE{0:1}}
    \|i when bp|            & \ABS &      &      \\
    \|post o|               & \UNK &      &      \\
    \|(post o) when not bp| & \UNK &      &      \\
    \|o|                    & \UNK &      &      \\
    \hline
    \hline
    \noalign{\gdef\CYCLE{0:2}}
    \|i when bp|            & \ABS & \ABS &      \\
    \|post o|               & \UNK & \UNK &      \\
    \|(post o) when not bp| & \UNK & \UNK &      \\
    \|o|                    & \UNK & \UNK &      \\
    \hline
    \hline
    \noalign{\gdef\CYCLE{0:3}}
    \|i when bp|            & \ABS & \ABS &    2 \\
    \|post o|               & \UNK & \UNK & \UNK \\
    \|(post o) when not bp| & \UNK & \UNK & \ABS \\
    \|o|                    & \UNK & \UNK &    2 \\
    \hline
    \|post o|               & \UNK & \CR2 & \UNK \\
    \|(post o) when not bp| & \UNK &    2 & \ABS \\
    \|o|                    & \UNK &    2 &    2 \\
    \hline
    \|post o|               & \CR2 &    2 & \UNK \\
    \|(post o) when not bp| &    2 &    2 & \ABS \\
    \|o|                    &    2 &    2 &    2 \\
    \hline
  \end{array}
  $$
  \end{minipage}
  \hfill
  \begin{minipage}[t]{.54\columnwidth}
  \scriptsize$$
  \begin{array}{|>{\SSEM}l<{\ENDSEM[\CYCLE]}|c@{\;}c@{\;}c@{\;}c@{\;}c@{\;}c|}
    \hline
    \noalign{\gdef\CYCLE{0:6}}
    \|i|                    &    0 &    1 &    2 &    3 &    4 &    5 \\
    \|bp|                   &   \F &   \F &   \T &   \T &   \F &   \T \\
    \hline
    \hline
    \noalign{\gdef\CYCLE{0:4}}
    \|i when bp|            & \ABS & \ABS &    2 &    3 &      &      \\
    \|post o|               &    2 &    2 & \UNK & \UNK &      &      \\
    \|(post o) when not bp| &    2 &    2 & \ABS & \ABS &      &      \\
    \|o|                    &    2 &    2 &    2 &    3 &      &      \\
    \hline
    \hline
    \noalign{\gdef\CYCLE{0:5}}
    \|i when bp|            & \ABS & \ABS &    2 &    3 & \ABS &      \\
    \|post o|               &    2 &    2 & \UNK & \UNK & \UNK &      \\
    \|(post o) when not bp| &    2 &    2 & \ABS & \ABS & \UNK &      \\
    \|o|                    &    2 &    2 &    2 &    3 & \UNK &      \\
    \hline
    \hline
    \noalign{\gdef\CYCLE{0:6}}
    \|i when bp|            & \ABS & \ABS &    2 &    3 & \ABS &    5 \\
    \|post o|               &    2 &    2 & \UNK & \UNK & \UNK & \UNK \\
    \|(post o) when not bp| &    2 &    2 & \ABS & \ABS & \UNK & \ABS \\
    \|o|                    &    2 &    2 &    2 &    3 & \UNK &    5 \\
    \hline
    \|post o|               &    2 &    2 &    3 & \UNK & \CR5 & \UNK \\
    \|(post o) when not bp| &    2 &    2 & \ABS & \ABS &    5 & \ABS \\  
    \|o|                    &    2 &    2 &    2 &    3 &    5 &    5 \\
    \hline
    \|post o|               &    2 &    2 &    3 & \CR5 &    5 & \UNK \\
    \hline
    \noalign{\vspace{2.5mm}}
    \hline
    \|backfill(i,bp)|
                            &    2 &    2 &    2 &    3 &    5 &    5 \\
    \hline
  \end{array}
  $$
  \end{minipage}
  \vspace{-3mm}
  \caption{Illustration of the synchronous semantics.}
  \label{fig:illus_sync}
  \vspace{-3mm}
\end{figure}


\subsection{Properties}
\label{sec:properties}

In the following we refer to streams with no unknown ($\UNK$) values as \emph{fully defined}.
One essential property of practically useful reactive programs is \emph{productivity}, a notion associated with (1) the ability to compute the least fixed point of a system of equation as the result of a finite computation, and (2) the resulting streams being \emph{fully defined}.
In the traditional Lustre setting, productivity (i.e.\ deadlock-freedom) stems from a conservative and modular interpretation of \emph{causality}: no \emph{instantaneous cycle} may exist when stepping through the computation of fixed points, i.e.\ when any recurrence on a given stream variable at index $n$ can be traced back to the previous index $n-1$ through the use of the \|fby| operator.
The presence of a \|fby| operator on all cycles in the dataflow graph defined by every node of the program is a sufficient condition for the absence of instantaneous cycles.
Weaker conditions and static analyses exist but none of these involve dependences on future indices, deemed non-reactive: the interested reader may refer to \cite{DBLP:journals/dafes/PouzetR10} for a survey and for extensions improving modularity and efficient compilation.

This pushes us to research an extended notion of productivity compatible with bidirectional reactions.
Intuitively, lookahead dependences may be allowed as long as they are eventually broken, i.e.\ they do not form an infinite recurrence into the future.
The problem is that allowing dependences into the future, even bounded ones, automatically gives rise to instantaneous cycles by composition with conventional dependences on the past.
Reasoning on dependences only is unlikely to lead to solution.
This intuition leads us to define the extended notion of \emph{bidirectional causality}.

\begin{definition}[Bidirectional causality]
  \label{def:causal}
  A program $P$ is bidirectionally causal if and only if the following condition hold on the synchronous interpretation $\SSEM P \ENDSEM : \VENV^d \to \VENV^{d'}$:
  $$\forall m \in \mathbb{N}, \exists z \in \mathbb{N}, \SSEM P \ENDSEM[m] = \SSEM P \ENDSEM[0:m][m+z]$$
\end{definition}

We are also interested in a stronger notion of \emph{bidirectional bounded reactivity}, analogous to the bounded memory, instantaneous termination property of synchronous reactive languages \cite{lustreRTSS}.
To this end, we combine bidirectional causality with a notion of termination suitable to stream functions.
Let $\MAP : \DSYNC \to \DSYNC$ be a Scott-continuous function; $\MAP$ \emph{terminates up to index} $n \in \mathbb{N}$ if and only if $\exists m \in \mathbb{N}, \STH{\FIX(\MAP)}{0:n} = \STH{\MAP^m(\UNK^\omega)}{0:n}$; and $\MAP$ \emph{terminates} if and only if it terminates up to index $n$ for all $n \in \mathbb{N}$.
A synchronous program $P$ terminates (resp.\ terminates up to index $n$) if and only if all fixed point computations terminate (resp.\ terminate up to index $n$).

\begin{definition}[Bidirectional bounded reactivity]
  \label{def:bounded}
  A program $P$ is bidirectionally bounded reactive if and only if the two following conditions hold on the synchronous interpretation $\SSEM P \ENDSEM : \VENV^d \to \VENV^{d'}$: (i) $\SSEM P \ENDSEM$ terminates, and (ii) $\exists \bar{z} \in \mathbb{N}, \forall m \in \mathbb{N}, \SSEM P \ENDSEM[m] = \SSEM P \ENDSEM[0:m][m+\bar{z}]$.
\end{definition}

Let us now state two theorems that are essential to making the interpretation of the semantics effective and practical.

\begin{theorem}[Finite and bounded memory interpretation]
  \label{thm:bounded}
  ~
  \begin{itemize}
  \item Finite memory (possibly unbounded): considering a bidirectionally causal program $P$ and the look-ahead offset $z$ guaranteed by bidirectional causality at a given index $m$, only needs to store windows of stream values of size $z+1$ to effectively interpret the semantics.
  \item Bounded memory: considering a bounded reactive program $P$ and the look-ahead offset $\bar{z}$ guaranteed by bounded reactivity for all indices, only needs to store windows of stream values of size $\bar{z}+1$ to effectively interpret the semantics.
  \end{itemize}
\end{theorem}

The proof stems from the observation that incremental left-to-right computation is possible for \|fby| operators, holding the value at the previous index only.

In the following, \emph{causal} is synonymous with \emph{bidirectionally causal}, and \emph{bounded reactive} is synonymous with \emph{bidirectionally bounded reactive}, unless stated otherwise.
It is immediate that a bounded reactive program is causal.

\begin{theorem}[Eventual progress]
  \label{thm:progress}
  For a causal program $P$, the synchronous interpretation up to a certain index coincides with the synchronous semantics with a finite look-ahead horizon applied to a bounded future of the program's inputs.
  For a bounded reactive program, a global bound exists for both look-ahead horizon and inputs.
  
  Formally, given program $P$ with $\SSEM P \ENDSEM : \VENV^d \to \VENV^{d'}$, if $P$ is causal then
  $$
  \forall \EI \in \DSYNC^d, \forall m\in\mathbb{N}, \exists z\in\mathbb{N},
  \SSEM P(\EI) \ENDSEM[0:m] = \SSEM P(\STH{\EI}{0:m+z} \CONS \UNK^\omega) \ENDSEM[0:m][m+z]
  $$
  and if $P$ is bounded reactive then
  $$
  \forall \EI \in \DSYNC^d, \exists \bar{z}\in\mathbb{N}, \forall m\in\mathbb{N},
  \SSEM P(\EI) \ENDSEM[0:m] = \SSEM P(\STH{\EI}{0:m+\bar{z}} \CONS \UNK^\omega) \ENDSEM[0:m][m+\bar{z}]
  $$
\end{theorem}

The proof is inductive on the structure of semantic equations.

\section{From denotational semantics to operational simulation}

The semantics of the previous section has the advantage introducing backwards recurrences in a conventional Lustre-like setting with minimal changes to the semantic domain.
To this end, it makes the simplifying assumption that programs are normalized, i.e. that constants are present at every cycle, that the \|fby| operator of forward recurrences is activated according to its first input, that the \|post| operator of backward recurrences is present at every cycle, and that the execution environment provides the status of every input variable at every cycle.
From these primary sources of control information, the status of every other variable can be incrementally constructed at every cycle by traversing every program equation to compute the status of the equation outputs based on that of the equation inputs.
In this process, every equation is traversed exactly once per cycle.

Formalizations of classical reactive languages rely on similar normalization.
In most cases, it is assumed that the activation of every variable and equation is described using a predicate, called a \emph{clock}, which can be provided by the programmer or computed by means of static analysis---the clock calculus.
The main downside of this classical approach is that the clocks need to be available and the normalization to take place even \emph{before one is able to define the program semantics}.
Indeed, the natural programming style implemented in classical compilers of dataflow reactive languages,\footnote{Such as \href{https://www-verimag.imag.fr/lustre-v6.html}{Lustre} or \href{https://gitlab.inria.fr/synchrone/heptagon}{Heptagon}.} does not require the definition of clocks nor does it expose the user with the need to normalize programs.
An additional drawback is that such normalization may actually be wasteful of memory resources, by forcing the allocation of memory for all inputs of a node at every cycle irrespectively of their downstream sampling through normalization.

In this section, we solve this conundrum by providing a semantics for programs that are not heavily normalized.
In doing this, we actually solve an old problem of reactive programming: introducing a semantic reference to establish the correctness of the static analysis algorithms producing clock annotations.

The semantics defined in this section has three more advantages:
\begin{itemize}
\item It provides support for the efficient simulation of bidirectionally causal programs (see \defref{causal}). This is actually the main reason to present this new normalization-free semantics in the context of bidirectional reactivity: bidirectionality within a given cycle is exactly what is needed to eliminate the need for normalization in the first place; while more general bidirectional reactivity amounts to extending the semantics and simulation to multi-cycle windows.
\item Specifically, one may simulate finite executions of bidirectionally bounded reactive programs (see \defref{bounded}) in a finite number of steps and with a finite memory footprint.
\item From an system and input/output standpoint, the semantics directly captures the behavior of programs that drive their inputs, possibly driven by their outputs, an important demand-driven pattern in both embedded and datacenter applications.
\end{itemize}
We formally prove that the simulator provided in \secref{simulation} implements the semantics of this section.
We also conjecture that our semantics is a strict generalization of that of \secref{semantics} and of that of classical reactive languages such as Lustre. Proving these results is work in progress.

\subsection{Domain definition}
\paragraph{Single-variable domain}
Given a variable $v$, we denote with $T_v$ the set of values of its
data type. We also overload $T_v$ to stand the data type itself. The
semantics may also assign to a variable, during a cycle, another 4
values with specific meaning:
\begin{itemize}
\item $\UNK$ represents the absence of any information on the
  status of the variable (at the current point in execution).
\item $\ABS$ represents the absent status of a variable during a
  cycle. It is needed to ensure by semantic means, without requiring
  preliminary static analysis, that a variable value produced in one
  execution cycle is always consumed within the same cycle.
\item $\VP$ is the \emph{synchronization status} of a variable for which we determined that
  it is present, but have not yet computed the value. It allows the
  on-demand triggering of the computation of constants and of
  recurrence outputs. It also allows triggering input acquisition
  actions, as opposed to always assuming that the status is provided
  by the environment. This value is a new addition compared to the sparsely indexed stream domain of the previous section.
\item $\EE$ signifies that the variable is involved in a {\em
  synchronization error} in the current cycle -- its present/absent
  status is inconsistent with that of other variables. It is required
  to ensure confluence when control can propagate
  backwards.\footnote{Which means that, for the same input, the
  sycnhronization error can be first detected, depending on the
  equation evaluation order, in different equations. The use of $\EE$
  allows propagating the error towards a unique end-state.}
\end{itemize}
The domain of $v$ is therefore $D_v= T_v\cup \{\UNK,\ABS,\VP,\EE\}$,
endowed with the partial order $\leq$ defined by: $\UNK$ and $e$
respectively are the infimum and supremum of $D_v$, and $\VP$ is
lower than the values $x\in T_v$. We denote the (always defined)
supremum operator of $D_v$ with $\vee$.

We define $nv:D_v\rightarrow \{\UNK,\ABS,\VP,\EE\}$ as the function that
strips the value of a variable, preserving only its synchronization status:
$$
  nv(x) = \left\{\begin{array}{ll}
          ? & \mbox{if \(x\in T_v\)} \\
          x & \mbox{otherwise}
          \end{array}\right.
$$

\paragraph{Multi-variable domain}
Given a set of variables $V$, we represent their semantic status at a
given execution step with elements of $D_V = \prod_{v\in V}D_v$. Given $v\in V$
and $c\in D_V$, we denote with $\VAL{c}{v}$ the value of $v$ in $c$. We
overload the $\UNK$ notation to (also) denote the element of $D_V$
with $\VAL{c}{v}=\UNK$ for all $v\in V$. We similarly overload in $D_V$ the
notations $\ABS$, $\VP$, and $\EE$.
$D_V$ is naturally endowed with the
product domain structure, where $\UNK$ and $\EE$ are respectively least
and greatest elements, $\leq$ is the partial order over $D_V$, and
$\vee$ is the (completely-defined) supremum operator.

If $s\in D_V$, $v_1,\ldots,v_n\in V$ and $x_i\in D_{v_i}$, then
$\ASSIGN{s}{v_1,\ldots,v_n}{}{x_1,\ldots,x_n}$ is the element of $D_V$
that equals $x_i$ on $v_i$ and $\VAL{s}{v}$ for all $v\not\in V$. We also
denote
$\AASSIGN{s}{v_1,\ldots,v_n}{x_1,\ldots,x_n}=s\vee\ASSIGN{s}{v_1,\ldots,v_n}{}{x_1,\ldots,x_n}$
the update operator that never loses information.

We denote with $cerr$ the endomorphism $cerr:D_V\rightarrow D_V$ that
returns $\EE$ if at least one component of its input is $\EE$, and returns
its input otherwise.

\paragraph{Domain of streams}
If $V$ is the set of variables of a program or set of equations $p$,
the we will use streams of $D_V^\infty$ to represent both (finite or
infinite) execution traces and the semantic states used in the
construction of the traces. Recall that if $s\in D_V^\infty$, then
$\STH{s}{n}\in D_V$ is the element of $s$ of index $n\geq 0$. In a
trace/semantic state $s\in D_V^\infty$, we will say that execution has
reached cycle $n\geq 0$ if there exists $m\geq n$ with
$\STH{s}{m}\not=\UNK$. For $s\in D_V^\infty$ we denote with $len(s)\in
\mathbb{N}\cup\{\infty\}$ its length.

The partial order $\leq$ on $D_V$ induces a partial order $\leq$ on
$D_V^\infty$ as follows: $s^1\leq s^2$ if, by definition,
$len(s^1)\leq len(s^2)$ and $\STH{s^1}{i}\leq\STH{s^2}{i}$ for all
$0\leq i< len(s^1)$. Endowed with $\leq$, $D_V^\infty$ is a complete
lattice with infimum $\epsilon$ and supremum $\EE^\infty$. The supremum
operator is denoted $\vee$.

\subsection{Semantic functions\label{sec:semafun}}
Given a set $X$, we denote with $id_X$ the identity function over $X$.
For instance, $id_{D_V^\infty}$ is the identity function over
$D_V^\infty$.

Given a domain $(D,\leq)$, and another set $A$, we also denote with
$\leq$ the poinwise partial order induced on $D^A$ (the set of
functions from $A$ to $D$) by the partial order of $D$. We
respectively denote with $\vee$ and $\wedge$ the supremum and infimum
operators induced by $\leq$.

Assume $V$ is the set of variables used by a program or set of
equations $p$. We will define the semantics of $p$ as a function
$\BISEM{p}\in(D_V^\infty)^{D_V^\infty}$. We must always have:
\begin{itemize}
\item $id_{D_V^\infty}\leq\BISEM{p}$, so that applying the semantics does
  not lose information present in the input.
\item $\BISEM{p}$ is idempotent\footnote{Or, equivalently,
$\BISEM{p}(s)$ is a fixed point of $\BISEM{p}$ for all $s$.} so that
  all the values that can be computed by $p$ given the input $s$ will
  be present in $\BISEM{p}(s)$, and applying $\BISEM{p}$ a second time
  builds no new values.
\end{itemize}

\subsubsection{Fixpoint iteration function}
Recall that given a directed-complete partial order (CPO)
$(D,\leq)$, a continuous function $f:D\rightarrow D$, and $x\in D$,
the least fixed point of $f$ greater than $x$, denoted $\BIFIX{f}{x}$,
exists and can be computed as:
\begin{equation*}
  \BIFIX{f}{x}=\lim_{n\rightarrow\infty}f^n(x)
\end{equation*}

To ensure that $\BISEM{p}(s)$ can be incrementally computed for every
$s$, we will require that it is defined as the least fixed point
greater than $s$ of another function
$\ITER{p}\in(D_V^\infty)^{D_V^\infty}$ called the {\em fixpoint
  iteration function} of $p$. We require that $\ITER{p}$ is continuous
and that $\ITER{p}\geq id_{D_V^\infty}$.  Then, for all $s$:
\begin{equation*}
\BISEM{p}(s)=\BIFIX{\ITER{p}}{s}=\lim_{n\rightarrow\infty}\ITER{p}^n(s)
\end{equation*}

\subsubsection{Semantics of stateless equations/programs}
If $p$ involves no recurrences (no \verb|fby| or \verb|post|
equation), then its semantics can be described using a {\em stateless
  iteration function} $\SITER{p}:D_V\rightarrow D_V$ that must be
continuous and have the property $\SITER{p}\geq id_{D_V}$. The
iteration function $\ITER{p}$ is then defined by
$\STH{\ITER{p}(s)}{n}=\SITER{p}(\STH{s}{n})$ for all $0\leq
n<len(s)$. This function is greater than $id_{D_V^\infty}$ and is
continuous, allowing the definition of $\BISEM{p}$.

\subsection{Composing semantic functions\label{sec:composing}}
The semantics of \REACT primitives will be provided directly in
\secref{primitives}. The semantics of program fragments including
multiple statements/equations, and that of nodes, is provided in this
section. It is obtained by composition from that of the primitives.

\subsubsection{Notations}
Given two sets of variables $V\subseteq W$ and $x\in D_W$, we denote
with $\RESTRICT{x}{V}$ the restriction of $x$ to $V$, which is an
element of $D_V$.  In the other sense, given $y\in D_V$, we denote
with $\EXTEND{y}{W}$ its extension to $W$, which equals $y$ over $V$
and $\bot$ elsewhere. The restriction and extension operators are both
extended pointwise to (respectively) elements of $D_W^\infty$ and
$D_V^\infty$. Thus, given $s\in D_W^\infty$, $\RESTRICT{s}{V}$ is an
element of $D_V^\infty$. Similarly, if $s\in D_V^\infty$ then
$\EXTEND{s}{W}$ belongs to $D_W^\infty$.

The extension operator is generalized to allow the embedding of
$(D_V^\infty)^{D_V^\infty}$ into $(D_W^\infty)^{D_W^\infty}$. If
$f:D_V^\infty\rightarrow D_V^\infty$, then
$\EXTEND{f}{W}:D_W^\infty\rightarrow D_W^\infty$ is defined by
$\EXTEND{f}{W}(s) = \EXTEND{f(\RESTRICT{s}{V})}{W}$.

\subsubsection{Semantics of composition\label{sec:composition}}
We can now define the semantics of composition. The semantics of each
equation or set of equations is defined over the variables it
involves. For instance, the semantics of ``\verb|x = merge c y z|'' is
defined over domain $D_{\{x,c,y,z\}}$. When building the semantics of
a code fragment (or node) containing an equation, we first use the
extension operator to lift the semantics of the individual equation
onto the variable set of the code fragment or node, which allows us to
define composition by means of upper bound:
\begin{center}
  $\BISEM{eq_1;\ldots;eq_n}(s) = \BIFIX{\ITER{eq_1;\ldots;eq_n}}{s}$
  \hspace*{1cm}
  $\ITER{eq_1;\ldots;eq_n} = \bigvee_{i=1}^{n}(\EXTEND{\ITER{eq_i}}{W})$
\end{center}
where $W$ is the set of variables taken as input or produced as output
by equations $eq_1,\ldots,eq_n$. Note that by applying the
Knaster-Tarski theorem we also obtain $\BISEM{eq_1;\ldots;eq_n}(s) =
\BIFIX{\bigvee_{i=1}^{n}(\EXTEND{\BISEM{eq_i}}{W})}{s}$.

\subsubsection{Semantics of nodes\label{sec:nodesema}}
Consider an \REACT node $n$. We assume that its equations are
$eq_1,\ldots,eq_n$, and that $I$, $O$, and $L$ are respectively the
(mutually disjoint) sets of input variables, output variables, and
local variables that are neither inputs nor outputs.  Then,
$\BISEM{eq_1;\ldots;eq_n}:D_{I\cup O\cup L}^\infty\rightarrow D_{I\cup O\cup L}^\infty$
and we define the semantics of node $n$ with functions $\BISEM{n},\ITER{n} : D_{I\cup O}^\infty\rightarrow D_{I\cup O}^\infty$:
\begin{center}
  $\BISEM{n}(s) = \BIFIX{\ITER{n}}{s}$\hspace*{1cm}
  $\ITER{n}(s)=\RESTRICT{\ITER{eq_1;\ldots;eq_n}(\EXTEND{s}{I\cup O\cup L})}{I\cup O}$ 
\end{center}

\subsection{The ``\texttt{fby back}'' primitive}
The semantics of this section will cover not only infinite
executions/traces of \REACT programs,\footnote{Like the semantics in
\secref{semantics} does.} but also {\em finite} ones. In this case,
backwards recurrences fully mirror, in time, the behavior of forward
recurrences, including their initialization, which for backwards
recurrences will happen in the {\em last} execution cycle of the
recurrence.

As the \verb|post| primitive does not allow providing an
initialization value (instead producing an output that is absent in
the last execution cycle), and for symmetry w.r.t. the forward
recurrences, we replace \verb|post| with a new primitive whose general
form is ``$x = y \; \|fby back| \; z;$'' where $y$ and $z$ are
expressions of the same type and $x$ is a stream variable. This
primitive requires that $x$, $y$, and $z$ are present in the same
cycles. We will denote with $t_0<t_1<t_2<\ldots$ the indices where
$x$, $y$ and $z$ are all present, and let
$l\in\mathbb{N}\cup\{\infty\}$ be the number of such cycles. Then,
the recurrence relation defined by the primitive is:
$$
  x_{t_n}=
    \left\{\begin{array}{lll}
          y_{t_n} & \mbox{if $n=l-1$} &\mbox{\color{gray} --- initialization}\\
          z_{t_{n+1}} & \mbox{otherwise} &\mbox{\color{gray} --- recurrence backwards in time} \\
    \end{array}\right.
$$
Notice that the initialization case of the definition does not apply
for infinite executions/traces.

The \verb|post| operator can always be expressed based on \verb|fby back|:
{\small\begin{center}
\lstinline|x = post y;|\hspace*{5mm} becomes \hspace*{5mm}\lstinline|x = (y fby back y) when (false fby back true);|
\end{center}}
The converse is not true, as it would require a means to identify
the last execution cycle of the equation, which \verb|post| and the
other primitives do not provide.

\subsection{Normalization}
In the general form of the recurrence primitives \verb|fby| and
``\verb|fby back|'', the initialization can be performed using any
expression. However, this general form can always be reduced to
the case where the initializer is a constant:
{\small\begin{center}
  \lstinline|y = i fby      x;|  \hspace*{5mm} becomes \hspace*{5mm}     \lstinline|x = if (true fby      false) then i else (k fby      x);|
  \lstinline|y = i fby back x;|  \hspace*{5mm} becomes \hspace*{5mm}     \lstinline|x = if (true fby back false) then i else (k fby back x);|
\end{center}}
\noindent where the constant $k$ is any value of the type of \verb|x|.

This transformation improves the orthogonality of our set of
primitives, by letting conditional control {\em inside the cycle} be
always handled by stateless conditional execution primitives, whereas
the recurrence primitives only focus on the recurrence itself, $k$
being the recurrence initializer, which corresponds to the initial
value of the accumulator in a \verb|fold_left| or \verb|fold_right|
iterator, which is not computed during iteration. 

The definition of the semantics of the primitives will assume that
this transformation has been performed, and that furthermore complex
expressions have been decomposed, so that:
\begin{itemize}
\item Every equation contains a single dataflow operator.
\item No variable appears twice in an equation.
\item Constants are only used as:
  \begin{itemize}
  \item Right-hand side of {\em constant equations} with a single variable as output.
  \item State initializers in \verb|fby| and ``\verb|fby back|''
    equations.
  \end{itemize}
\end{itemize}

\subsection{Semantics of primitives\label{sec:primitives}}

\begin{definition}[Function application semantics]
  \label{def:funappsema}
  Consider a function $f:\prod_{i=1}^n T_{x_i}\rightarrow\prod_{j=1}^m T_{y_j}$,
  with $n\geq 1$ and the equation
  $eq=$``$\mathtt{(y}_1\mathtt{,}\ldots\mathtt{,y}_m\mathtt{)=f(x}_1\mathtt{,}\ldots\mathtt{,x}_n\mathtt{)}$''.
  Its stateless iterator function is
  $\SITER{eq}:\prod_{i=1}^nD_{x_i}\times\prod_{j=1}^m D_{y_j}
              \rightarrow
              \prod_{i=1}^nD_{x_i}\times\prod_{j=1}^m D_{y_j}$
  defined for any $v_i\in D_{x_i}$ and $w_j\in D_{y_j}$ by:
  \newline\noindent$\SITER{eq}(v_1,\ldots,v_n,w_1,\ldots,w_m)=$
  $$
  =
  \left\{\begin{array}{ll}
          cerr((v_1,\ldots,v_n,u_1\vee w_1,\ldots,u_m\vee w_m)) & \mbox{if $s=\VP$ and $\forall i:v_i\in T_{x_i}$} \\
          (s\vee v_1,\ldots,s\vee v_n,s\vee w_1,\ldots,s\vee w_m) & \mbox{otherwise}
  \end{array}\right.
  $$ where $s=nv(v_1)\vee\ldots\vee nv(v_n)\vee nv(w_1)\vee\ldots\vee
  nv(w_m)$, and (when $\forall i:v_i\in T_{x_i}$)
  $f(v_1,\ldots,v_n)=(u_1,\ldots,u_m)$.
\end{definition}
Notice that computation values are only propagated from inputs to the
outputs (in the first case of the definition, through the application
of $f$), whereas synchronization information will traverse the
equation in both senses (the second case of the definition, and the
application of $cerr$).

Whenever a synchronization error is identified on one of the variables,
it is propagated to all variables, either explicitly through the application
of $cerr$, or implicitly through the computation of $s$.

The semantics of constant equations (with a constant right-hand side)
can be provided as a sub-case of function application. However, for
clarity, we prefer defining it separately.
\begin{definition}[Constant semantics]
  \label{def:funappsema}
  Consider equation $eq=$``$\mathtt{y=k}$'', where $k\in T_y$.
  Its stateless iterator function is
  $\SITER{eq}:D_y\rightarrow D_y$ defined by:
  $$
  \SITER{eq}(w)
  =
  \left\{\begin{array}{ll}
          k\vee w & \mbox{if $w\geq\VP$} \\
          w & \mbox{otherwise}
  \end{array}\right.
  $$
\end{definition}
Notice that the value of the constant is produced when the status of
the output variable is ``$\VP$''. The same demand-driven protocol also
allows inputs to be read on-demand.

\begin{definition}[Semantics of \texttt{when}]\label{def:when}
  Consider the equation $eq=$``$\|$o$ = $y$ when $c$|$''.
  Its stateless iterator function is
  $\SITER{eq}:D_{\{o,y,c\}}\rightarrow D_{\{o,y,c\}}$ defined as:
  $$\SITER{eq} = cerr\circ when^o\circ when^y \circ when^c$$
  where:
  {\small
  \newline\noindent\hspace*{5mm}$
    when^c(s)
    =
    \left\{\begin{array}{ll}
          s\vee\ABS & \mbox{if $\VAL{s}{c} = \ABS$} \\
          \AASSIGN{s}{y}{\VP}         & \mbox{if $\VAL{s}{c} = \VP$} \\
          \AASSIGN{s}{o,y}{\VAL{s}{y}\vee\VP,\VP} & \mbox{if $\VAL{s}{c} = \T$} \\
          \AASSIGN{s}{o,y}{\ABS,\VP} & \mbox{if $\VAL{s}{c} = \F$} \\
          s               & \mbox{otherwise}
    \end{array}\right.
  $
  \newline\noindent\hspace*{5mm}$
    when^y(s)
    =
    \left\{\begin{array}{ll}
          s\vee\ABS & \mbox{if $\VAL{s}{y} = \ABS$} \\
          \AASSIGN{s}{c}{\VP}         & \mbox{if $\VAL{s}{y} \in T_y\cup\{\VP\}$} \\
          s               & \mbox{otherwise}
    \end{array}\right.
    \mspace{10mu}
    when^o(s)
    =
    \left\{\begin{array}{ll}
          s\vee\VP        & \mbox{if $\VAL{s}{o} \in T_y\cup\{\VP\}$} \\
          s               & \mbox{otherwise}
    \end{array}\right.
  $}
\end{definition}
Notice how the definition of $\SITER{eq}$ has been decomposed into
multiple semantic functions, each one representing the semantic
consequences of setting a particular input or output variable to a
particular value. For instance, in the definition of function
$when^c$, having $c=\T$ in the input state results in $y$ and $o$
being set to present, and if $y$ already has a value of $T_y$,
it propagates towards $o$. This propagation action can result in
$o$ taking the value or, if $o$ already has a value of $\ABS$
or $\EE$, its value becomes $\EE$.

To obtain the semantics of the primitive, the per-variable functions
are composed, and $cerr$ is applied so that an error detected in one
variable is propagated to all other variables.

\begin{definition}[Semantics of \texttt{merge}]
  Consider the equation $eq=$``$\|$o$ = merge $c$ $y$ $z$|$''.
  Its stateless iteration function is
  $\SITER{eq}:D_{\{o,c,y,z\}}\rightarrow D_{\{o,c,y,z\}}$
  defined as:
  $$\SITER{eq} = cerr\circ merce^o\circ merge^{y,z} \circ merge^c$$
  where:
  {\small
  \newline\noindent\hspace*{5mm}$
    merge^c(s)
    =
    \left\{\begin{array}{ll}
          s\vee\ABS & \mbox{if $\VAL{s}{c} = \ABS$} \\
          \AASSIGN{s}{o}{\VP}         & \mbox{if $\VAL{s}{c} = \VP$} \\
          \AASSIGN{s}{o,y,z}{\VAL{s}{y}\vee\VP,\VP,\ABS} & \mbox{if $\VAL{s}{c} = \T$} \\
          \AASSIGN{s}{o,y,z}{\VAL{s}{z}\vee\VP,\ABS,\VP} & \mbox{if $\VAL{s}{c} = \F$} \\
          s               & \mbox{otherwise}
    \end{array}\right.
  $
  \newline\noindent\hspace*{5mm}$
    merge^{y,z}(s)
    =
    \left\{\begin{array}{ll}
          \AASSIGN{s}{c}{\VP} & \mbox{if \{$\VAL{s}{y},\VAL{s}{z}\} \cap( T_y\cup\{\VP\})\not=\emptyset$} \\
          s               & \mbox{otherwise}
    \end{array}\right.
    \mspace{10mu}
    merge^o(s)
    =
    \left\{\begin{array}{ll}
          s\vee\ABS  & \mbox{if $\VAL{s}{o} =\ABS$} \\
          \AASSIGN{s}{c}{\VP}  & \mbox{if $\VAL{s}{o} \in T_y\cup\{\VP\}$} \\
          s              & \mbox{otherwise}
    \end{array}\right.
  $}
\end{definition}
Notice that the semantics of \verb|merge| requires that the
Boolean condition value is received before the data input
can be propagated towards the output.

\begin{definition}[Semantics of \texttt{fby}]
  Consider the equation $eq=$``$\|$o$ = k fby $y$|$'', where $k\in T_o=T_y$.
  Its iteration function is $\ITER{eq}:D_{\{o,y\}}^\infty\rightarrow D_{\{o,y\}}^\infty$
  defined by $\ITER{eq} = fby^{val}\circ fby^{sync}$, where:
  {\small
    \newline\noindent\hspace*{5mm}
    $\STH{fby^{sync}(s)}{n} = \AASSIGN{\STH{s}{n}}{o,y}{nv(\VAL{\STH{s}{n}}{y}),nv(\VAL{\STH{s}{n}}{o})}$
                             {\color{gray}---synchronization, involving no recurrence}
    \newline\noindent\hspace*{5mm}
    $\STH{fby^{val}(s)}{n}  = \AASSIGN{\STH{s}{n}}{o}{curr^{fw}(s,n)}$    
                             {\color{gray}---forward recurrence}
  }
  \newline\noindent with:
  {\small\newline\noindent\hspace*{5mm}$
    curr^{fw}(s,n)
    =
    \left\{\begin{array}{ll}
       \VAL{\STH{s}{n}}{o}                         & \mbox{if $\VAL{\STH{s}{n}}{o}\not\in T_o\cup\{\VP\}$} \\
       k\vee\VAL{\STH{s}{n}}{o}                    & \mbox{if $\VAL{\STH{s}{n}}{o}\in T_o\cup\{\VP\}$ and $\forall k<n:\VAL{\STH{s}{k}}{o}=\ABS$} \\
       \VAL{\STH{s}{m}}{y}\vee\VAL{\STH{s}{n}}{o}  & \mbox{if $\VAL{\STH{s}{n}}{o},\VAL{\STH{s}{m}}{y}\in T_o\cup\{\VP\}$ and $\forall m<k<n:\VAL{\STH{s}{k}}{o}=\ABS$} \\
       \bot    & \mbox{otherwise}
    \end{array}\right.
  $}
\end{definition}
\begin{definition}[Semantics of ``\texttt{fby back}'']\label{def:fbyback}
  Consider the equation $eq=$``$\|$o$ = k fby back $y$|$'', where $k\in T_o=T_y$.
  Its iteration function is $\ITER{eq}:D_{\{o,y\}}^\infty\rightarrow D_{\{o,y\}}^\infty$
  defined by $\ITER{eq} = fbyback^{val}\circ fby^{sync}$, where:
  {\small
    \newline\noindent\hspace*{5mm}
    $\STH{fbyback^{val}(s)}{n}  = \AASSIGN{\STH{s}{n}}{o}{curr^{bw}(s,n)}$    
                             {\color{gray}---backward recurrence}
  }
  \newline\noindent with:
  {\small\newline\noindent\hspace*{5mm}$
    curr^{bw}(s,n)
    =
    \left\{\begin{array}{ll}
       \VAL{\STH{s}{n}}{o}                         & \mbox{if $\VAL{\STH{s}{n}}{o}\not\in T_o\cup\{\VP\}$} \\
       k\vee\VAL{\STH{s}{n}}{o}                    & \mbox{if $\VAL{\STH{s}{n}}{o}\in T_o\cup\{\VP\}$ and $\forall n<k<=len(s)-1:\VAL{\STH{s}{k}}{o}=\ABS$} \\
       \VAL{\STH{s}{m}}{y}\vee\VAL{\STH{s}{n}}{o}  & \mbox{if $\VAL{\STH{s}{n}}{o},\VAL{\STH{s}{m}}{y}\in T_o\cup\{\VP\}$ and $\forall n<k<m:\VAL{\STH{s}{k}}{o}=\ABS$} \\
       \bot    & \mbox{otherwise}
    \end{array}\right.
  $}
\end{definition}
Note that, if the equation
$eq=$``$\|$o$ = $x$ fby back $y$|$''
is activated (inputs or outputs present) only in a finite number of cycles
of an infinite trace, then in the last activation cycle its output
remains at value $\bot$.
Indeed, determining that the equation is inactive in all subsequent
cycles cannot be done algorithmically over an infinite number of
cycles. Such $\bot$ values can be filtered out using \verb|when|
equations.

The last primitive we need to consider is node instantiation. 
\begin{definition}[Node instantiation semantics]
  \label{def:funappsema}
  Consider equation $eq=$``$\mathtt{(y}_1\mathtt{,}\ldots\mathtt{,y}_m\mathtt{)=n(x}_1\mathtt{,}\ldots\mathtt{,x}_n\mathtt{)}$'',
  where $\mathtt{n}$ is a node (not a function). Then $\ITER{eq}=\ITER{n}$ (as defined in \secref{nodesema}).
\end{definition}
Note that this definition makes the semantics of instantiation
equations depend on that of the instantiated nodes, whereas
\secref{nodesema} makes the semantics of nodes depend on that of
equations. The correction of this recursive definition is guaranteed
here by the rule governing module instantiation: no node can directly
or indirectly instantiate itself. This rule implies that the nodes
instantiated (directly or indirectly) by a node $n$ form a {\em
  finite} tree with $n$ as root, each node having as children the node
instances it has as equations.

\subsection{Fixed point computation\label{sec:simulation}}
Consider a node $p$. Assume its set of variables is $W$ and its set of inputs is $I\subseteq W$.
Consider a valuation of its inputs $i\in D_I^\infty$. Then, the semantics of $p$ for input $i$
is $\BISEM{p}(\II)$, where $\II=\EXTEND{i}{V}$.

Under the definition of \secref{simulation}, $\BISEM{p}(\II)$ is
obtained by cyclically applying $\ITER{p}$ to compute
$\ITER{p}^n(\II)$. But one application of $\ITER{p}$ is already an
infinite process, as it may build values at an infinity of cycles of
$\II$. Therefore, this organization of the fixpoint definition cannot
be used as a pattern for operational implementation.

To allow implementation, we need to choose a different order for the
computation of variable values. We choose to always compute new values
at the lowest cycle where it is possible. This is also the
implementation strategy of classical reactive languages, where values
are built cycle by cycle. In these languages, when all values of cycle
$n$ are constructed, execution can move to cycle $n+1$. But when
recurrences backwards in time are allowed, this is not always
possible. For instance, in the training of RNNs by means of BPTT (as
in \figref{lstmbackprop}) a full recurrent input sample is processed
by the forward path before back-propagation can start. For this
reason, multiple cycles can can have values under construction.

At every step of the fixpoint computation, we denote with $lst$ the
last cycle under construction. Whenever no new values can be built
between cycles $0$ and $lst$ without using information of cycles of index
$n>lst$, we increment $lst$.

We denote with $todo_n(s)$ the set of all equations that, in
fixed point simulation state $s$, would change the value of a
variable in cycle $n$. It is formally defined as:
$$todo_n(s) = \{eq\mid \STH{\ITER{eq}(\RESTRICT{s}{V_{eq}})}{n} > \STH{(\RESTRICT{s}{V_{eq}})}{n} \}$$
where $V_{eq}$ denotes the set of variables of equation $eq$.  Given $s\in D_V^\infty$ and
$x\in D_V$, we denote $s[n\leftarrow x]$ the element of $D_V^\infty$
that equals $x$ in cycle $n$ and $\STH{s}{m}$ in every cycle
$m\not=n$.

With these notations, the computation of the fixed point is realized
using the algorithm in \figref{fixpoint}. To simplify presentation,
the algorithm assumes that node $p$ instantiates no other nodes. To
cover node instantiation, the simulation state must be generalized to
include not only the valuation of variables, but also the simulation
state of every instantiated node, hierarchically under the form of a finite tree, and the state computation
and update (lines 13-14) must be updated accordingly.

\begin{figure}
  \begin{center}
{\small$$
  \begin{array}{ll}
_1  & \hspace{0cm}s \mathtt{\ :=\ } \bot[0\leftarrow \STH{\II}{0}]\ \color{gray}\mathtt{(*\ }\text{initial state}\mathtt{\ *)} \\
_2  & \hspace{0cm}lst \mathtt{\ :=\ } 0\ \color{gray}\mathtt{(*\ }\text{initial lst}\mathtt{\ *)} \\
_3  & \hspace{0cm}\mathbf{while(true)}\ \color{gray}\mathtt{(*\ }\text{iteration,\ a\ priori\ infinite}\mathtt{\ *)}
\\
_4  & \hspace{1cm}  n = 0\ \color{gray}\mathtt{(*\ }\text{start\ computation\ of\ the\ first\ cycle\ with\ actions\ to\ execute}\mathtt{\ *)}\\
_5  & \hspace{1cm}  \mathbf{while\ } todo_{n}(s)=\emptyset \mathbf{\ do}\ \color{gray}\mathtt{(*\ }\text{in cycle n no equation can be executed}\mathtt{\ *)}  \\
_6  & \hspace{2cm}    \mathbf{if\ } n<len(\II) \mathbf{\ then\ }
                         n \mathtt{\ :=\ } n + 1\ \color{gray}\mathtt{(*\ }\text{try the next cycle}\mathtt{\ *)}
                         \color{black}\mathbf{\ else\ return} \\
_7  & \hspace{2cm}    \mathbf{if\ } n > lst \mathbf{\ then\ }\color{gray}\mathtt{(*\ }\text{if\ needed\ open\ cycles\ to\ simulation}\mathtt{\ *)} \\
_8  & \hspace{3cm}      lst \mathtt{\ :=\ } n\ \color{gray}\mathtt{(*\ }\text{in cycle n I can execute no equation, try the next cycle}\mathtt{\ *)} \\
_8  & \hspace{3cm}      s := s[lst\leftarrow \STH{\II}{n}]\ \color{gray}\mathtt{(*\ }\text{include\ the\ inputs\ of\ the\ newly\ open\ cycle\ in\ the\ state}\mathtt{\ *)} \\
_{10}& \hspace{2cm}    \mathbf{end} \\
_{11}& \hspace{1cm}  \mathbf{end} \\
_{12}& \hspace{1cm}  eq \mathtt{\ :=\ } choose(todo_{n})\ \color{gray}\mathtt{(*\ }\text{choose\ equation\ to\ execute from non-void set}\mathtt{\ *)}\\
_{13}& \hspace{1cm}  act \mathtt{\ :=\ } \STH{\ITER{eq}(\RESTRICT{s}{V_{eq}})}{n}\ \color{gray}\mathtt{(*\ }\text{compute\ update\ action}\mathtt{\ *)}\\
_{14}& \hspace{1cm}  s \mathtt{\ :=\ } s[n\leftarrow \STH{s}{n}\vee (\EXTEND{act}{W})]\ \color{gray}\mathtt{(*\ }\text{change\ simulation\ state}\mathtt{\ *)} \\
_{15}& \hspace{0cm}\mathbf{end}
\end{array}
$$}
  \caption{Fixed point iteration algorithm. It only terminates (in line 6) if the input is finite.\label{fig:fixpoint}}
  \end{center}
\end{figure}

Notice that the fixed point iteration algorithm considers inputs
incrementally. This means that state representations used during
iteration are always finite, allowing implementation. 

\subsubsection{Equivalence with denotational semantics}
Along a given execution of the algorithm in \figref{fixpoint},
we shall denote with $\AITER{p}{m}(\II)$ the value of $s$ at the
end of the $m^{th}$ iteration of the main iteration loop. Then,
the fixed point iteration algorithm and the semantics are related
by the following result:
\begin{theorem}\label{thm:DenotVsSimul}
  Under the previous definitions, for every $n\geq 0$, if
  $\STH{\BISEM{p}(\II)}{n:}\not=\STH{i}{n:}$, then there exists
  $m\geq 0$ such that $\STH{\BISEM{p}(\II)}{0:n}=$ $\STH{\AITER{p}{m}(\II)}{0:n}$
\end{theorem}
Assuming in the theorem hypothesis that
$\STH{\BISEM{P}(\II)}{n:}\not=\STH{i}{n:}$ amounts to assuming
that the semantics assigns at least one new value to at least one
variable in a cycle of index $k\geq n$. Without this assumption,
the fixed point iteration algorithm would remain stuck in the
internal while loop that computes $n$ (lines 7-13), as the
semantics has no values to compute.
\begin{proof}[Proof of \thmref{DenotVsSimul}]
  We first prove that $\AITER{p}{m}\BISEM{p}(\II)\leq \BISEM{p}(\II)$ for all $m$.
  This is done by induction. For $m=0$,
  $\AITER{p}{0}(\II)=\bot[0\leftarrow \STH{\II}{0}]\leq \II\leq\BISEM{p}(\II)$.
  For the induction step, we denote with $f^{m+1}$ the function used to
  construct $\AITER{p}{m+1}(\II)$ from $\AITER{p}{m}(\II)$, as defined in
  lines 15-16 of the algorithm. As $f^{m+1}$ is obtained from $\ITER{eq^{m+1}}$ for
  some equation $eq^{m+1}$, it is by construction smaller than
  $\ITER{p}$ (itself the union of $\ITER{eq}$ for all equations $eq$). Then:
  $\AITER{p}{m+1}(\II)=f^{m+1}(\AITER{p}{m}(\II))\leq\ITER{p}(\AITER{p}{m}(\II))\leq
  \BISEM{p}(\AITER{p}{m}(\II))\leq\BISEM{p}(\BISEM{p}(\II))$.
  And since $\BISEM{p}(\II)$ is a fixed point of $\BISEM{p}$, the
  induction step is complete.

  From $\AITER{p}{m}(\II)\leq \BISEM{p}(\II)$ for all $m$ we can then
  deduce $\STH{\AITER{p}{m}(\II)}{0:n}\leq\STH{\BISEM{p}(\II)}{0:n}$
  for all $m,n$.
 
  To complete the proof of our theorem, we need to find a value of $m$
  for which the reverse inequality holds. Since
  $\BISEM{p}(\II)=\lim_{r\rightarrow\infty}\ITER{p}^r(\II)$, we obtain
  that there exists $r_0$ such that
  $\STH{\BISEM{p}(\II)}{0:n}=\STH{\ITER{p}^r(\II)}{0:n}$ for all $r\geq
  r_0$. Then, proving our theorem can be done by proving that for every
  $r\geq 0$ and $n\geq -1$ there exists $m_{r,n}$ such that
  \begin{equation}\label{eq:toprove}
    \STH{\AITER{p}{m_{r,n}}(\II)}{0:n}\geq\STH{\ITER{p}^r(\II)}{0:n}
  \end{equation}
  We prove this by outer induction over $r$ and inner induction over
  $n$. Recall that $\STH{x}{0:-1}$ is the empty stream, so that
  property \eqref{eq:toprove} is trivially satisfied for $n=-1$ for any
  $r$ and $m_{r,-1}$. We choose $m_{r,-1}=0$ for all $r$, and thus have
  solved the base case for the inner induction for all $r$.

  \paragraph{Induction over $r$, base case.}
  For $r=0$ and $n\geq -1$, we need to prove that for every $n$ there
  exists $m_{0,n}$ such that
  $\STH{\AITER{p}{m_0}(\II)}{0:n}\geq\STH{\II}{0:n}$. For $n=-1$ the
  property has been established above. For $n\geq 0$, from the theorem
  hypothesis, $\STH{\BISEM{P}(\II)}{n:}\not=\STH{\II}{n:}$, which implies
  that the fixed point computation proceeds to at least cycle $n$, and
  each time $lst$ is incremented the values of $\II$ for cycle $lst$ are
  included in $s$ (in line 11).  Thus, for $m_{0,n}=n$,
  $\STH{\AITER{p}{m_{0,n}}(\II)}{0:n}\geq\STH{\II}{0:n}$.

  \paragraph{Induction step over $r$.}
  As induction hypothesis we assume that for all $n$ we
  have constructed $m_{r,n}$ such that
  $\STH{\AITER{p}{m_{r,n}}(\II)}{0:n}\geq\STH{\ITER{p}^r(\II)}{0:n}$.
  From the monotony of $\ITER{p}$ we have
  $\ITER{p}^r(\II)\leq\ITER{p}^{r+1}(\II)$.
  
  If $\STH{\ITER{p}^r(\II)}{0:n}=\STH{\ITER{p}^{r+1}(\II)}{0:n}$, then
  from the induction hypothesis we have
  $\STH{\AITER{p}{m_r}(\II)}{0:n}\geq\STH{\ITER{p}^{r+1}(\II)}{0:n}$, so
  the induction step is verified for $m_{r+1,n}=m_{r,n}$.

  If $\STH{\ITER{p}^r(\II)}{0:n}\not=\STH{\ITER{p}^{r+1}(\II)}{0:n}$, then
  $\STH{\ITER{p}^r(\II)}{0:n}<\STH{\ITER{p}^{r+1}(\II)}{0:n}$.  This is
  equivalent to there existing $u\geq 1$ distinct variables
  $v_1,\ldots,v_u\in W$ and $0\leq k_1,\ldots,k_u\leq n$ 
  such that
  $\STH{\ITER{p}^{r+1}(\II)}{k_l}[v_l]>\STH{\ITER{p}^r(\II)}{k_l}[v_l]$.
  We will determine that for every $1\leq l\leq u$ there exists
  $m^l>m_{r,n}$ such that $\STH{\ITER{p}^{r+1}(\II)}{k_l}[v_l]=\STH{\AITER{p}{m^l}(\II)}{k_l}[v_l]$.
  Once this is proven, property \eqref{eq:toprove} is satisfied for $m_{r+1,n}=\max_{l=1}^u m^l$.

  Consider $(v_l,k_l)$ one of the $u$ pairs with
  $\STH{\ITER{p}^{r+1}(\II)}{k_l}[v_l]>\STH{\ITER{p}^r(\II)}{k_l}[v_l]$.
  By construction, $v_l$ cannot be an input of $p$, but an output of
  a unique equation of $p$. Let $eq$ be this equation.
  From the definition of $\ITER{p}$, we know that the value is
  produced by $\ITER{eq}$ in the context provided
  by $\ITER{p}^r(\II)$. If $eq$ is not a ``\verb|fby back|'' equation,
  then this context can be restrained to $\STH{\ITER{p}^r(\II)}{0:n}$,
  which is included in $\STH{\AITER{p}{m_{r,n}}(\II)}{0:n}$. Equation $eq$
  is therefore part of $todo_{k_l}$ in state $\AITER{p}{m_{r,n}}(\II)$. As such,
  it will be executed in a finite number of steps of the fixed point
  iteration (in the worst case, after all variables are assigned a
  new value in all cycles $c\leq k_l$). Then, we can set $m^l=m_{r,n}+\mid W\mid*k_l$,
  where $\mid W\mid$ is the cardinal of $W$.

  Finally, the case where $eq=$``$\|$v_l$ = $x$ fby back $y$|$''.
  From
  $\STH{\ITER{p}^{r+1}(\II)}{k_l}[v_l]>\STH{\ITER{p}^r(\II)}{k_l}[v_l]$ we
  determine that $\STH{\ITER{p}^{r+1}(\II)}{k_l}[v_l]$ is constructed
  during the $(r+1)^{th}$ application of $\ITER{p}$. Constructing this
  value is done according to \defref{fbyback}. Then, the construction of
  $\STH{\ITER{p}^{r+1}(\II)}{k_l}[v_l]$ depends on a finite number of
  absent or present variable values of $\ITER{p}^r(\II)$. Let $N$ be
  the maximum cycle index of such a variable value. Then, we know that
  all variable values needed to construct $\STH{\ITER{p}^{r+1}(\II)}{k_l}[v_l]$
  are present in $\AITER{p}{m_{r,N}}(\II)$. Then, we can set
  $m^l=m_{r,n}+\mid W\mid*N$. The proof is completed.
\end{proof}

\section{Implementation}
To allow the evaluation of \REACT, we have implemented a full
front-end and simulator named \texttt{mlrc}. It covers the language
introduced in the paper, extended with NumPy-like functional arrays
and a (data) type system featuring 3 scalar types (\verb|bool|,
\verb|int|, and a type foating point type \verb|num| that must be used
on all variables subject to back-propagation) and a classical notation
for tensor shapes. Thus, ``\verb|i:num [2,3]|'' defines variable
\verb|i| as a 2-dimensional floating point tensor with 2 lines and 3
columns. To avoid ambiguity, scalars are represented as tensors with
0 dimensions. Thus, ``\verb|c:bool []|'' defines a Boolean scalar.
Our \texttt{mlrc} compiler relies on \href{https://www.tensorflow.org/xla}{XLA} to allow efficient
execution of functions such as \texttt{Conv2D}.

The front-end includes type and shape inference phases that will assign
a fixed type and shape to all variables, starting from those of the
inputs. This is required to allow simulation. 

Prior to simulation, the front-end will perform full type/shape
inference, inlining of all nodes in the main node, and a pass of
advanced constant folding (which itself requires clock inference) to
reduce the state size.

The algorithm in \figref{fixpoint} provides a first method to execute
\REACT programs, one that was designed for generality (equivalence with
denotational semantics) and simplicity. To attain efficiency, it must
be further refined. The overall event-driven simulation method can be
largely improved by optimizing the computation of the $todo_n$ sets
and the search for an equation to activate, at each step of the
algorithm. In \texttt{mlrc} we do both. Each time a new value is
computed for a variable $v$ in cycle $n$, we determine which equations
having $v$ as input can compute a new value. These equations are then
added to the corresponding $todo_k$ set, to be removed when the
corresponding value is computed. We also exclude from the search in
lines 4-11 the {\em completed} cycles where all values have already
been computed. Furthermore, the valuations of these cycles can be
deleted from the simulation state provided the cycle is not adjacent
to a non-completed one. This allows reducing to a minimum the number
of currently open cycles of the simulation state.

Of course, in an ML context most performance gains require replacing
the event-driven simulation with statically-scheduled code optimizable
by a compiler such as XLA. We currently only do it for individual
functions.

\section{Implementation}
\label{sec:implementation}

Considering a \emph{synchronous, bidirectionally causal} program $P$, the properties established in the previous section provide a concrete path to its interpretation.
For the computation up to a given index $m$, \thmref{progress} states that it is sufficient to evaluate the semantic equations up to some bounded index $m+z$, and \thmref{bounded} states that $z+1$ live indices are sufficient to interpret the semantics equations of $P$ up to index $m$.
The value of $z$ is a function of $m$ in general, unless the program is actually bounded reactivewhere one may further establish an upper bound $\bar{z}$ such that a maximal amount of $\bar{z}+1$ live indices are sufficient.

This is not yet sufficient to allow a practical implementation: one still needs to compute $z$ for any $m$.
To that end, let $\WORKLIST_n$ denote the set of equations whose right-hand-side lead to a new (higher) value of a variable at cycle $n$:
$\WORKLIST_m = \{(x = e) \in P \mid \STH{\SSEM x = e \ENDSEM}{m} (\VENV) \GTS \STH{\VENV(x)}{m} \}$.
This predicate may be evaluated recursively from the synchronous transformers in \figref{sync_transformers}, considering only the instantaneous left-hand side of the $\CONS$ operator, and memoizing the \emph{window of live indices} lower or higher than the cycle $m$ of interest.
In particular, one may choose to always compute new values at the lowest index possible, matching the implementation of classical reactive languages where values are computed cycle by cycle.
When all values at index $m$ have reached a fixed point, execution can move to cycle $m+1$, and memory for cycle $m$ can be reclaimed except for values held by \|fby| operators.
When the program is bounded reactive, the window of live indices remains bounded, making the implementation practical.

We implemented a front-end and interpreter for \REACT called \texttt{mlrc}. The syntax presented in the paper is extended with NumPy-like functional arrays and a (data) type system featuring 3 scalar types (\verb|bool|, \verb|int| and a floating point type \verb|num| that must be used on all variables subject to backpropagation). Functional arrays feature a classical notation for tensor shapes: e.g.\ ``\verb|i:num [2,3]|'' defines variable \verb|i| as a 2-dimensional floating point tensor with 2 rows and 3 columns. Scalars are represented as tensors with 0 dimensions: e.g.\ ``\verb|c:bool []|'' defines a Boolean scalar.
The front-end includes type and shape inference passes to assign a type and shape to all variables, starting from the inputs.
The version \texttt{mlrc} provided as supplementary material only supports primitive Ocaml functions and operators, while the full version relies on XLA \cite{XLA} to support realistic neural networks.

While the algorithm we just sketched interprets any \REACT program, there is a lot of room for performance improvement. In particular, the \texttt{mlrc} implementation specializes the computation of $\WORKLIST_n$ sets and the search for the next equation to evaluate. Each time a new value is computed for a variable $v$ at cycle $n$, one may determine which equations having $v$ as input can compute a new value. These equations are then added to the corresponding $\WORKLIST_k$ set, and retired when the value is eventually computed. One may also exclude from the search the completed cycles where all values have already been computed, and one may then recollect the storage locations associated with these unless the cycle is adjacent to a non-completed one.

\section{Conclusion}

We introduced \REACT, a bidirectional reactive programming language on unbounded streams with both forward and backward dependences.
We illustrated the modularity and expressiveness of the language on a variety of ML models, algorithms and usage scenarios, with unprecedented parsimony---four traditional reactive constructs plus one original backward recurrence primitive.
We provided an denotational synchronous semantics with higher expressiveness than traditional synchronous Kahn networks preserving its main productivity and bounded memory execution properties.
Starting from a single reactive program, one may transform it structurally to compute both forward and reverse-mode gradients, leveraging bidirectional reactive semantics on streams instead of referring to an external data structure such as a gradient tape or cache.
From this single reactive program, one may also derive backpropagation and (batched) training algorithms by inserting dedicated control signals around initialization and sampling primitives.
The language also unifies the expression of models interacting with their environment, from reinforcement learning to front-end/back-end data processing and real-time control.
A prototype is provided as supplementary material, as well as detailed simulations and discussion of the denotational semantics.

\newpage

\bibliographystyle{plain}
\bibliography{paper}

\appendix

\section{Kahn semantics}

For reference and comparison, we recall the definition of the Kahn domain then tentatively derive an asynchronous semantics \cite{Kahn74}.
We show by example that such a semantics, being limited to prefix-closed stream values, leads to non-interesting least fixed points in the context of bidirectional reactive programming.

\paragraph{Kahn domain}
As announced earlier, we first experiment with providing \REACT with an asynchronous dataflow semantics over finite and infinite streams by exending the Kahn network formalism.

Let $\leq_K$ be the prefix order over streams: $x \leq_K y \iff \exists z, y = x \CONS z$.
The ordered set $\DKAHN = (\DATATYPE^\infty, \leq_K ,\varepsilon)$---known as the \emph{Kahn domain} \cite{Kahn74}---is a CPO with infimum $\varepsilon$ and no supremum.

\paragraph{Asynchronous semantics on the Kahn domain}
A tentative Kahn semantics extended to bidirectional reactivity is given in \figref{kahn_semantics}.
The interpretation of the associated stream transformers is given in \figref{kahn_transformers}.
The mark $\KF{}$ distinguishes the syntactic construct from its interpretation as a stream transformer.

Notice the \|when $\F$| case may shrink an infinite stream into a finite (possibly empty) one.
Notice also $\KSEM \|post|\ e \ENDSEM = \KSEM e\ \|when (false fby true)| \ENDSEM$: in this asynchronous semantics, \|post| happens to be syntactic sugar for sampling with the stream $\F \CONS \T^\omega$.

\begin{figure}[h!tb]
\begin{eqnarr}
\KSEM c \ENDSEM (\VENV) & \equaldef & c^\omega \\
\KSEM x \ENDSEM (\VENV) & \equaldef & \VENV(x) \\
\KSEM \|$i$ fby $e$| \ENDSEM (\VENV) & \equaldef & \SF{\|fby|} (\VENV(\KSEM i \ENDSEM (\VENV)), \VENV(\KSEM e \ENDSEM (\VENV))) \\
\KSEM \|post $e$| \ENDSEM (\VENV) & \equaldef & \SF{\|post|} (\VENV(\KSEM e \ENDSEM (\VENV))) \\
\KSEM \|$e$ when $b$| \ENDSEM (\VENV) & \equaldef & \SF{\|when|} (\VENV(\KSEM e \ENDSEM (\VENV), \KSEM b \ENDSEM (\VENV))) \\
\KSEM \|merge $b$|\ \ET\ \EF\, \ENDSEM (\VENV) & \equaldef & \SF{\|merge|} (\VENV(\KSEM b \ENDSEM (\VENV), \KSEM \ET \ENDSEM (\VENV), \KSEM \EF\, \ENDSEM (\VENV)) \\
\KSEM \|$f$($e$)| \ENDSEM (\VENV) & \equaldef & \VENV(f) (\KSEM e \ENDSEM) \\
\KSEM \|node $f$($i$)->($o$)|\ \textit{eq} \ENDSEM (\VENV) & \equaldef & f \BIND \lambda d. \KSEM \pi_o(\!\textit{eq}) \ENDSEM (\VENV, i \BIND d) \textrm{ where } i \subseteq \FV(\!\textit{eq}) \land o \subseteq \DV(\!\textit{eq}) \\
\KSEM \|$x$ = $e$| \ENDSEM (\VENV) & \equaldef & \KSEM e \ENDSEM (\VENV, x \BIND x_\FIX) \textrm{ where } x_\FIX = \FIX (\lambda d. \KSEM e \ENDSEM (\VENV, x \BIND d))
\end{eqnarr}
\caption{Kahn semantics.}
\label{fig:kahn_semantics}

\begin{eqnarr}
\KF{\|fby|} (\EI, \varepsilon) & \equaldef & \varepsilon \\
\KF{\|fby|} (\varepsilon, \ES) & \equaldef & \varepsilon \\
\KF{\|fby|} (\EI, \ES) & \equaldef & \SHD{\EI} \CONS \ES \\
\KF{\|post|} (\varepsilon) & \equaldef & \varepsilon \\
\KF{\|post|} (\SHD{\ES} \CONS \STH{\ES}{1:}) & \equaldef & \STH{\ES}{1:} \\
\KF{\|when|} (\ES, \varepsilon) & \equaldef & \varepsilon \\
\KF{\|when|} (\varepsilon, \EB) & \equaldef & \varepsilon \\
\KF{\|when|} (\SHD{\ES} \CONS \STH{\ES}{1:}, \T \CONS \STH{\EB}{1:}) & \equaldef & \SHD{\ES} \CONS \KF{\|when|} (\STH{\ES}{1:}, \STH{\EB}{1:}) \\
\KF{\|when|} (\SHD{\ES} \CONS \STH{\ES}{1:}, \F \CONS \STH{\EB}{1:}) & \equaldef & \KF{\|when|} (\!\STH{\ES}{1:}, \STH{\EB}{1:}) \\
\KF{\|merge|} (\varepsilon, \ET, \EF) & \equaldef & \varepsilon \\
\KF{\|merge|} (\T \CONS \STH{\EB}{1:}, \varepsilon, \EF) & \equaldef & \varepsilon \\
\KF{\|merge|} (\F \CONS \STH{\EB}{1:}, \ET, \varepsilon) & \equaldef & \varepsilon \\
\KF{\|merge|} (\T \CONS \STH{\EB}{1:}, \SHD{\ET} \CONS \STH{\ET}{1:}, \EF) & \equaldef &
\SHD{\ET} \CONS \KF{\|merge|} (\!\STH{\EB}{1:}, \STH{\ET}{1:}, \EF) \\
\KF{\|merge|} (\F \CONS \STH{\EB}{1:}, \ET, \SHD{\EF} \CONS \STH{\EF}{1:}) & \equaldef &
\SHD{\EF} \CONS \KF{\|merge|} (\!\STH{\EB}{1:}, \ET, \STH{\EF}{1:}) \end{eqnarr}
\caption{Stream transformers for the primitives of the Kahn semantics.}
\label{fig:kahn_transformers}
\end{figure}

\figref{illus_kahn} illustrates these definitions on the \|backfill| example.
Each row corresponds to the evaluation of one rule in the fixed point computation, effectively computing a new finite or infinite stream one row at a time.
Blank cells are meant for visual alignment only, they are not meaningful in the $\DKAHN$ domain.
From the $\KF{\|merge|} (\F \CONS \STH{\EB}{1:}, \ET, \varepsilon)$ case of the $\KF{\|merge|}$ stream transformer, the Kahn semantics computes an empty stream for \|o| at the first iteration.
As a result, it immediately reaches a fixed point, making no progress at all in the computation of \|o|.
This deadlocking behavior is the direct result of the domain operating on prefixed-closed streams and is common to reactive languages on streams \cite{lustreRTSS,ElliottHudak97:Fran}.

\begin{figure}[h!tb]
  \begin{lstlisting}[basicstyle=\ttfamily\scriptsize]
node backfill(i,bp)->(o)
  o = merge bp (i when bp) ((post o) when not bp);
  \end{lstlisting}
  \vspace{-.7\baselineskip}
  {\scriptsize$$
  \begin{array}{|>{\KSEM}l<{\ENDSEM}|ccccccc|}
    \hline
    \|i|                      &    0 &    1 &    2 &    3 &    4 &    5 & \ldots \\
    \|bp|                     &   \F &   \F &   \T &   \T &   \F &   \T & \ldots \\
    \hline
    \hline
    \|i when bp|              &      &      &    2 &    3 &      &    5 & \ldots \\
    \|post o|                 &      &      &      &      &      &      & \\
    \|(post o) when not bp|   &      &      &      &      &      &      & \\
    \|o|                      &      &      &      &      &      &      & \\
    \hline
    \hline
    \|backfill (i, bp)| &      &      &      &      &      &      & \\
    \hline
  \end{array}$$}
  \caption{Illustration of the Kahn semantics deadlocking on a backward recurrence.}
  \label{fig:illus_kahn}
\end{figure}

\section{Cycle-by-cycle semantic simulations}

Beyond the \|backfill| example discussed in the main paper, \figref{illus_sync_sample} ilustrates a more complex case of sampling over values computed with the lookahead operator.
Notice how $\UNK$ values in \|o| stick in at cycles 0 and 1, and eventually disappear at cycle 2 ($\SSEM \|o| \ENDSEM[0:3]$).

\figref{illus_sync_merge} explores the semantics of merge (starting at cycle 2 for readability): the interpretation switches from $\UNK$ to known ({\CB blue}) values much earlier, without waiting for the prefix of \|c| to be known in full.
\figref{illus_sync_rec_merge} goes one step further, illustrating the expressiveness of the synchronous semantics by revisiting the recurrent merge case (starting at cycle 2).

\begin{figure}[h!tb]
\begin{lstlisting}[basicstyle=\ttfamily\scriptsize]
node sample_backfill(i,bp)->(o)
  nh = false fby (not nh);
  c = backfill(nh,bp);
  o = i when c;
\end{lstlisting}
  \vspace{-.7\baselineskip}
  {\scriptsize$$
  \begin{array}{|>{\SSEM}l<{\ENDSEM[\CYCLE]}|cccccc|}
    \hline
    \noalign{\gdef\CYCLE{0:6}}
    \|i|                    &    0 &    1 &    2 &    3 &    4 &    5 \\
    \|bp|                   &   \F &   \F &   \T &   \T &   \F &   \T \\
    \|nh|                   &   \F &   \T &   \F &   \T &   \F &   \T \\
    \hline
    \hline
    \noalign{\gdef\CYCLE{0:1}}
    \|nh when bp|           & \ABS &      &      &      &      &      \\
    \|post c|               & \UNK &      &      &      &      &      \\
    \|(post c) when not bp| & \UNK &      &      &      &      &      \\
    \|c|                    & \UNK &      &      &      &      &      \\
    \|o|                    & \UNK &      &      &      &      &      \\
    \hline
    \hline
    \noalign{\gdef\CYCLE{0:2}}
    \|nh when bp|           & \ABS & \ABS &      &      &      &      \\
    \|post c|               & \UNK & \UNK &      &      &      &      \\
    \|(post c) when not bp| & \UNK & \UNK &      &      &      &      \\
    \|c|                    & \UNK & \UNK &      &      &      &      \\
    \|o|                    & \UNK & \UNK &      &      &      &      \\
    \hline
    \hline
    \noalign{\gdef\CYCLE{0:3}}
    \|nh when bp|           & \ABS & \ABS &   \F &      &      &      \\
    \|post c|               & \UNK & \UNK & \UNK &      &      &      \\
    \|(post c) when not bp| & \UNK & \UNK & \ABS &      &      &      \\
    \|c|                    & \UNK & \UNK &   \F &      &      &      \\
    \|o|                    & \UNK & \UNK & \ABS &      &      &      \\
    \hline
    \|post c|               & \UNK &\CR\F & \UNK &      &      &      \\
    \|(post c) when not bp| & \UNK &   \F & \ABS &      &      &      \\
    \|c|                    & \UNK &   \F &   \F &      &      &      \\
    \|o|                    & \UNK & \ABS & \ABS &      &      &      \\
    \hline
    \|post c|               &\CR\F &   \F & \UNK &      &      &      \\
    \|(post c) when not bp| &   \F &   \F & \ABS &      &      &      \\
    \|c|                    &   \F &   \F &   \F &      &      &      \\
    \|o|                    & \ABS & \ABS & \ABS &      &      &      \\
    \hline
    \hline
    \noalign{\gdef\CYCLE{0:4}}
    \|nh when bp|           & \ABS & \ABS &   \F &   \T &      &      \\
    \|post c|               &   \F &   \F & \UNK & \UNK &      &      \\
    \|(post c) when not bp| &   \F &   \F & \ABS & \ABS &      &      \\
    \|c|                    &   \F &   \F &   \F &   \T &      &      \\
    \|o|                    & \ABS & \ABS & \ABS &    3 &      &      \\
    \hline
    \hline
    \noalign{\gdef\CYCLE{0:5}}
    \|nh when bp|           & \ABS & \ABS &   \F &   \T & \ABS &      \\
    \|post c|               &   \F &   \F & \UNK & \UNK & \UNK &      \\
    \|(post c) when not bp| &   \F &   \F & \ABS & \ABS & \UNK &      \\
    \|c|                    &   \F &   \F &   \F &   \T & \UNK &      \\
    \|o|                    & \ABS & \ABS & \ABS &    3 & \UNK &      \\
    \hline
    \hline
    \noalign{\gdef\CYCLE{0:6}}
    \|nh when bp|           & \ABS & \ABS &   \F &   \T & \ABS &   \T \\
    \|post c|               &   \F &   \F & \UNK & \UNK & \UNK & \UNK \\
    \|(post c) when not bp| &   \F &   \F & \ABS & \ABS & \UNK & \ABS \\
    \|c|                    &   \F &   \F &   \F &   \T & \UNK &   \T \\
    \|o|                    & \ABS & \ABS & \ABS &    3 & \UNK &    5 \\
    \hline
    \|post c|               &   \F &   \F &   \T & \UNK &\CR\T & \UNK \\
    \|(post c) when not bp| &   \F &   \F & \ABS & \ABS &   \T & \ABS \\  
    \|c|                    &   \F &   \F &   \T &   \T &   \T &   \T \\
    \|o|                    & \ABS & \ABS & \ABS &    3 &    4 &    5 \\
    \hline
    \hline
    \|sample_backfill(i,bp)|
                            & \ABS & \ABS & \ABS &    3 &    4 &    5 \\
    \hline
  \end{array}
  $$}
  \caption{Illustration of the synchronous semantics, sampling on a lookahead condition.}
  \label{fig:illus_sync_sample}
\end{figure}

\begin{figure}[h!bt]
  \begin{lstlisting}[basicstyle=\ttfamily\scriptsize]
node merge_backfill(i,bp)->(o)
  nh = false fby (not nh);
  c = backfill(nh,bp);
  o = merge c (i when c) (-i when not c);
  \end{lstlisting}
  \vspace{-.7\baselineskip}
  {\scriptsize$$
  \begin{array}{|>{\SSEM}l<{\ENDSEM[\CYCLE]}|cccccc|}
    \hline
    \noalign{\gdef\CYCLE{0:6}}
    \|i|                    &    0 &    1 &    2 &    3 &    4 &    5 \\
    \|bp|                   &   \F &   \F &   \T &   \T &   \F &   \T \\
    \|nh|                   &   \F &   \T &   \F &   \T &   \F &   \T \\
    \hline
    \hline
    \noalign{\gdef\CYCLE{0:3}}
    \|nh when bp|           & \ABS & \ABS &   \F &      &      &      \\
    \|post c|               & \UNK & \UNK & \UNK &      &      &      \\
    \|(post c) when not bp| & \UNK & \UNK & \ABS &      &      &      \\
    \|c|                    & \UNK & \UNK &   \F &      &      &      \\
    \|i when c|             & \UNK & \UNK & \ABS &      &      &      \\
    \|-i when not c|        & \UNK & \UNK &   -2 &      &      &      \\
    \|o|                    & \UNK & \UNK &\CB-2 &      &      &      \\
    \hline
    \|post c|               & \UNK &\CR\F & \UNK &      &      &      \\
    \|(post c) when not bp| & \UNK &   \F & \ABS &      &      &      \\
    \|c|                    & \UNK &   \F &   \F &      &      &      \\
    \|i when c|             & \UNK & \ABS & \ABS &      &      &      \\
    \|-i when not c|        & \UNK &   -1 &   -2 &      &      &      \\
    \|o|                    & \UNK &\CB-1 &   -2 &      &      &      \\
    \hline
    \|post c|               &\CR\F &   \F & \UNK &      &      &      \\
    \|(post c) when not bp| &   \F &   \F & \ABS &      &      &      \\
    \|c|                    &   \F &   \F &   \F &      &      &      \\
    \|i when c|             & \ABS & \ABS & \ABS &      &      &      \\
    \|-i when not c|        &    0 &   -1 &   -2 &      &      &      \\
    \|o|                    & \CB0 &   -1 &   -2 &      &      &      \\
    \hline
    \hline
    \noalign{\gdef\CYCLE{0:4}}
    \|nh when bp|           & \ABS & \ABS &   \F &   \T &      &      \\
    \|post c|               &   \F &   \F & \UNK & \UNK &      &      \\
    \|(post c) when not bp| &   \F &   \F & \ABS & \ABS &      &      \\
    \|c|                    &   \F &   \F &   \F &   \T &      &      \\
    \|i when c|             & \ABS & \ABS & \ABS &   3  &      &      \\
    \|-i when not c|        &    0 &   -1 &   -2 & \ABS &      &      \\
    \|o|                    &    0 &   -1 &   -2 &   3  &      &      \\
    \hline
    \|post c|               &   \F &   \F &\CR\T & \UNK &      &      \\
    \hline
    \hline
    \noalign{\gdef\CYCLE{0:5}}
    \|nh when bp|           & \ABS & \ABS &   \F &   \T & \ABS &      \\
    \|post c|               &   \F &   \F &   \T & \UNK & \UNK &      \\
    \|(post c) when not bp| &   \F &   \F & \ABS & \ABS & \UNK &      \\
    \|c|                    &   \F &   \F &   \F &   \T & \UNK &      \\
    \|i when c|             & \ABS & \ABS & \ABS &   3  & \UNK &      \\
    \|-i when not c|        &    0 &   -1 &   -2 & \ABS & \UNK &      \\
    \|o|                    &    0 &   -1 &   -2 &   3  & \UNK &      \\
    \hline
    \hline
    \noalign{\gdef\CYCLE{0:6}}
    \|nh when bp|           & \ABS & \ABS &   \F &   \T & \ABS &   \T \\
    \|post c|               &   \F &   \F &   \T & \UNK & \UNK & \UNK \\
    \|(post c) when not bp| &   \F &   \F & \ABS & \ABS & \UNK & \ABS \\
    \|c|                    &   \F &   \F &   \F &   \T & \UNK &   \T \\
    \|i when c|             & \ABS & \ABS & \ABS &   3  & \UNK &   5  \\
    \|-i when not c|        &    0 &   -1 &   -2 & \ABS & \UNK & \ABS \\
    \|o|                    &    0 &   -1 &   -2 &    3 & \UNK & \CB5 \\
    \hline
    \|post c|               &   \F &   \F &   \T & \UNK &\CR\T & \UNK \\
    \|(post c) when not bp| &   \F &   \F & \ABS & \ABS &   \T & \ABS \\  
    \|c|                    &   \F &   \F &   \F &   \T &   \T &   \T \\
    \|i when c|             & \ABS & \ABS & \ABS &   3  &    4 &   5  \\
    \|-i when not c|        &    0 &   -1 &   -2 & \ABS & \ABS & \ABS \\
    \|o|                    &    0 &   -1 &   -2 &    3 &    4 &    5 \\
    \hline
    \|post c|               &   \F &   \F &   \T &\CR\T &   \T & \UNK \\
    \hline
    \hline
    \|merge_backfill(i,bp)|
                            &    0 &   -1 &   -2 &    3 &    4 &    5 \\
    \hline
  \end{array}$$}
  \caption{Illustration of the synchronous semantics, merging on a lookahead condition.}
  \label{fig:illus_sync_merge}
\end{figure}

\begin{figure}[h!bt]
  \begin{lstlisting}[basicstyle=\ttfamily\scriptsize]
node rec_merge_backfill(bp)->(o)
  nh = false fby (not nh);
  c = merge bp (nh when bp) ((post o) when not bp);
  o = backfill(nh,c);
  \end{lstlisting}
  \vspace{-.7\baselineskip}
  {\scriptsize$$
  \begin{array}{|>{\SSEM}l<{\ENDSEM[\CYCLE]}|cccccc|}
    \hline
    \noalign{\gdef\CYCLE{0:6}}
    \|nh|                   &   \F &   \T &   \F &   \T &   \F &   \T \\
    \|bp|                   &   \F &   \F &   \T &   \T &   \F &   \T \\
    \hline
    \hline
    \noalign{\gdef\CYCLE{0:3}}
    \|nh when bp|           & \ABS & \ABS &   \F &      &      &      \\
    \|post o|               & \UNK & \UNK & \UNK &      &      &      \\
    \|(post o) when not bp| & \UNK & \UNK & \ABS &      &      &      \\
    \|c|                    & \UNK & \UNK &   \F &      &      &      \\
    \|nh when c|            & \UNK & \UNK & \ABS &      &      &      \\
    \|(post o) when not c|  & \UNK & \UNK & \UNK &      &      &      \\
    \|o|                    & \UNK & \UNK & \UNK &      &      &      \\
    \hline
    \hline
    \noalign{\gdef\CYCLE{0:4}}
    \|nh when bp|           & \ABS & \ABS &   \F &   \T &      &      \\
    \|post o|               & \UNK & \UNK & \UNK & \UNK &      &      \\
    \|(post o) when not bp| & \UNK & \UNK & \ABS & \ABS &      &      \\
    \|c|                    & \UNK & \UNK &   \F &   \T &      &      \\
    \|nh when c|            & \UNK & \UNK & \ABS &   \T &      &      \\
    \|(post o) when not c|  & \UNK & \UNK & \UNK & \ABS &      &      \\
    \|o|                    & \UNK & \UNK & \UNK &   \T &      &      \\
    \hline
    \|post o|               & \UNK & \UNK &\CR\T & \UNK &      &      \\
    \|(post o) when not bp| & \UNK & \UNK & \ABS & \ABS &      &      \\
    \|c|                    & \UNK & \UNK &   \F &   \T &      &      \\
    \|nh when c|            & \UNK & \UNK & \ABS &   \T &      &      \\
    \|(post o) when not c|  & \UNK & \UNK &   \T & \ABS &      &      \\
    \|o|                    & \UNK & \UNK &   \T &   \T &      &      \\
    \hline
    \|post o|               & \UNK &\CR\T &   \T & \UNK &      &      \\
    \|(post o) when not bp| & \UNK &   \T & \ABS & \ABS &      &      \\
    \|c|                    & \UNK &   \T &   \F &   \T &      &      \\
    \|nh when c|            & \UNK &   \T & \ABS &   \T &      &      \\
    \|(post o) when not c|  & \UNK & \ABS &   \T & \ABS &      &      \\
    \|o|                    & \UNK &   \T &   \T &   \T &      &      \\
    \hline
    \|post o|               &\CR\T &   \T &   \T & \UNK &      &      \\
    \|(post o) when not bp| &   \T &   \T & \ABS & \ABS &      &      \\
    \|c|                    &   \T &   \T &   \F &   \T &      &      \\
    \|nh when c|            &   \F &   \T & \ABS &   \T &      &      \\
    \|(post o) when not c|  & \ABS & \ABS &   \T & \ABS &      &      \\
    \|o|                    &   \F &   \T &   \T &   \T &      &      \\
    \hline
    \hline
    \noalign{\gdef\CYCLE{0:5}}
    \|nh when bp|           & \ABS & \ABS &   \F &   \T & \ABS &      \\
    \|post o|               &\CR\T &   \T &   \T & \UNK & \UNK &      \\
    \|(post o) when not bp| &   \T &   \T & \ABS & \ABS & \UNK &      \\
    \|c|                    &   \T &   \T &   \F &   \T & \UNK &      \\
    \|nh when c|            &   \F &   \T & \ABS &   \T & \UNK &      \\
    \|(post o) when not c|  & \ABS & \ABS &   \T & \ABS & \UNK &      \\
    \|o|                    &   \F &   \T &   \T &   \T & \UNK &      \\
    \hline
    \hline
    \noalign{\gdef\CYCLE{0:6}}
    \|nh when bp|           & \ABS & \ABS &   \F &   \T & \ABS &   \T \\
    \|post o|               &   \T &   \T &   \T & \UNK & \UNK & \UNK \\
    \|(post o) when not bp| &   \T &   \T & \ABS & \ABS & \UNK & \UNK \\
    \|c|                    &   \T &   \T &   \F &   \T & \UNK &   \T \\
    \|nh when c|            &   \F &   \T & \ABS &   \T & \UNK &   \T \\
    \|(post o) when not c|  & \ABS & \ABS &   \T & \ABS & \UNK & \UNK \\
    \|o|                    &   \F &   \T &   \T &   \T & \UNK &   \T \\
    \hline
    \|post o|               &   \T &   \T &   \T & \UNK &\CR\T & \UNK \\
    \|(post o) when not bp| &   \T &   \T & \ABS & \ABS &   \T & \UNK \\
    \|c|                    &   \T &   \T &   \F &   \T &   \T &   \T \\
    \|nh when c|            &   \F &   \T & \ABS &   \T &   \F &   \T \\
    \|(post o) when not c|  & \ABS & \ABS &   \T & \ABS & \ABS & \UNK \\
    \|o|                    &   \F &   \T &   \T &   \T &   \F &   \T \\
    \hline
    \|post o|               &   \T &   \T &   \T &\CR\F &   \T & \UNK \\
    \hline
    \hline
    \|rec_merge_backfill(i,bp)|
                            &   \F &   \T &   \T &   \T &   \F &   \T \\
    \hline
  \end{array}$$}
  \caption{Illustration of the synchronous semantics, merging on a recurrently defined lookahead condition.}
  \label{fig:illus_sync_rec_merge}
\end{figure}

\end{document}